\documentstyle[PASJadd,psfig]{PASJ95}
\begin{document}
\title{X-Ray Probing of the Central Regions of Clusters of Galaxies} 

\author{Kazuo {\sc Makishima},$^{1,2}$  Hajime {\sc Ezawa},$^3$  Yasushi {\sc Fukazawa},$^4$ \\ 
       Hirohiko {\sc Honda},$^5$ Yasushi {\sc Ikebe},$^6$ 
Tsuneyoshi {\sc Kamae},$^{4}$  Ken'ich {\sc Kikuchi},$^7$\\
Kyoko {\sc Matsushita},$^6$ Kazuhiro {\sc Nakazawa},$^1$ Takaya {\sc Ohashi},$^8$\\
Tadayuki {\sc Takahashi},$^5$ Takayuki {\sc Tamura},$^9$ and Haiguang {\sc Xu}$^{10}$
\vspace*{1mm}
\\
{\small\it 
$^1$ Department of Physics, University of Tokyo, 7-3-1 Hongo, Bunkyo-ku,Tokyo 113-0033}\\
{\small\it E-mail (KM) maxima@phys.s.u-tokyo.ac.jp} \\       
\vspace{2mm}
{\small\it $^2$ Research Center for the Early Universe (RESCEU), 
   University of Tokyo, 7-3-1 Hongo, Bunkyo-ku,Tokyo 113-0033}\\
\vspace{2mm}
{\small\it $^3$ Nobeyama Radio Observatory, National Astronomical Observatory,
Minamimaki, Minamisaku, Nagano 384-1305}\\ 
\vspace{2mm}
{\small\it $^4$ Department of Physical Science, Hiroshima University,
1-3-1 Kagamiyama, Higashi-hiroshima 739-8526}\\ 
\vspace{2mm}
{\small\it $^5$ Institute of Space and Astronautical Science, 3-1-1 Yoshinodai, Sagamihara, Kan
agawa 226-8510}\\ 
\vspace{2mm}
{\small\it $^6$ Max-Planck-Institut f\"{u}r Extraterrestrische Physik, Postfach 1603, D-85740, 
Garching, Germany} \\
\vspace{2mm}
{\small\it $^7$ Space Utilization Research Program, National Space Development Agency of Japan,
Tsukuba, Ibaraki 305-8505}\\ 
\vspace{2mm}
{\small\it $^8$ Department of Physics, Tokyo Metropolitan University, Hachioji, Tokyo 192-0397}\\  
\vspace{2mm}
{\small\it $^9$ SRON, Sorbonnelaan 2, 3584 CA Utrecht, the Netherlands} \\
\vspace{2mm}
{\small\it $^{10}$ Institute for Space Astrophysics, Shanghai Jiao Tong University,
Huashan Road 1954, Shanghai 200030, PRC} \\
}
\abst{
\vspace*{2mm}
Results of ASCA X-ray study of central regions of
medium-richness clusters of galaxies are summarized,
emphasizing differences between cD and non-cD clusters.
The intra-cluster medium (ICM) is likely to consist of
two (hot and cool) phases within $\sim 100$ kpc of a cD galaxy,
where the ICM metallicity is also enhanced.
In contrast, the ICM in non-cD clusters appears to be isothermal
with little metallicity gradient right to the center.
The gravitational potential exhibits a hierarchical nesting around cD galaxies,
while a total mass-density profile with a central cusp
is indicated for a non-cD cluster Abell~1060.
The iron-mass-to-light ratio of the ICM decreases
toward the center in both types of clusters,
although it is radially constant in peripheral regions.
The silicon-to-iron abundance ratio in the ICM
increases with the cluster richness,
but remains close to the solar ratio around cD galaxies.
These overall results are interpreted
without appealing to the popular cooling-flow hypothesis.
Instead, an emphasis is put on the  
halo-in-halo structure formed around cD galaxies.
}
\kword{galaxies: clustering --- interstellar medium --- X-rays: spectra}
\maketitle
\thispagestyle{headings}

%=====================
\section{INTRODUCTION}
%=====================
Central regions of galaxy clusters are of particular interest
with respect to the formation and evolution of galaxies and clusters.
At or near the center of some clusters, 
we often find a single dominant elliptical galaxy (cD galaxy).
With large extended stellar envelopes
(e.g. Schombert 1986; Johnstone et al. 1991),
these objects form by far the most luminous class of galaxies.
Extensive theoretical and numerical studies generally suggest 
that they were formed relatively early in the course of cluster evolution 
(e.g., Merritt 1985; Dubinski 1998; Ghigna et al. 2000),
but detailed comparison with the observation is yet to be performed.
In other clusters, 
we find several giant elliptical galaxies instead of a cD galaxy.
The origin of the difference between clusters with and without 
cD galaxies is not yet clear.

As noticed by Jones, Forman (1984),
the presence of a cD galaxy in a cluster also affects
the properties of the intra-cluster medium (ICM), 
which fills the intra-cluster space and emit intense X-rays (Sarazin 1988).
Around a cD galaxy, 
the soft X-ray surface brightness often exhibits a strong excess 
above a $\beta$ model fitted to the outer-region brightness profile
(e.g. Jones, Forman 1984; Edge et al. 1992);
we may call the phenomenon ``central excess emission'' (CEE). 
In addition, the X-ray emission from such an environment  
often involves a spectroscopic cool component,
hereafter called central cool component (CCC),
emitted from plasmas with a representative temperature of  $T \sim 1$ keV
(e.g. Fabian 1994; Fabian et al. 1994; Matsumoto et al. 1996).
So far, the CEE that is a spatial effect
and the CCC that is a spectroscopic one have been interpreted 
as different aspects of the same phenomenon, 
i.e., radiative cooling process and associated continuous ICM in-flow,
termed cooling flow (CF; e.g. Fabian 1994), 
proceeding in the densest part of the ICM.

The CF scenario was motivated by the short radiative 
cooling times of the ICM inferred from X-ray observations,
and reinforced by the CCC and CEE phenomena.
Theoretical treatment of CFs has been developed to such a degree 
of sophistication as inhomogeneous CFs (Nulsen 1986; Fabian 1994) 
and isothermal CFs (Nulsen 1998).
However, the CF hypothesis is still subject to several basic problems.
The heat conduction must somehow be suppressed
in order for a CF to develop (Takahara, Takahara 1979).
The fate of the large amount of cooled gas is controversial:
the possible X-ray absorbing gas (e.g. White et al 1991) may not be massive 
enough to act as a full sink for CFs (Allen, Fabian 1997),
and the presence of such an absorber itself is 
sometimes questioned (Sarazin 1997; Huang, Sarazin 1998; McNamara et al. 2000).
The searches for neutral hydrogen (e.g., Dwarakanath et al. 1994)
and coronal emission lines (e.g., Yan, Cohen 1995) have been unsuccessful.
Furthermore, numerical calculations of the cluster evolution predict too 
high X-ray luminosities when the ICM cooling is fully taken into account
(Suginohara, Ostriker 1998).
Although Cen, Ostriker (1999) predict 
that a major fraction of baryons in the universe is in the form
of warm gas of temperature $10^{5-7}$ K,
this component is predicted to diffusely distribute in the 
inter-galactic space rather than forming a cooling condensation.

X-ray observations by ASCA (Tanaka et al. 1994),
with its imaging capability in the previously unexplored energy range up to 10 keV
and its high spectroscopic performance,
have renewed our understanding of the CCC phenomenon.
Although we can clearly resolve the CCC in the ASCA spectra
taken from central regions of many cD clusters (e.g. Fabian et al. 1994),
the strongest CCC has been found to account for only 
a minor fraction of the X-ray emission therein (subsections 2.1 and 2.2).
This implies that the effect of ICM cooling 
was considerably overestimated previously (subsection 2.5).
Furthermore, the ICM around cD galaxies often 
exhibits a marked metallicity increase
and characteristic chemical compositions (subsection 4.3),
neither of which fits into the CF hypothesis.
Accordingly, we can more naturally interpret the CCC as an X-ray emitting 
inter-stellar medium (ISM) associated with the cD galaxy (subsection 2.6),
rather than as the cooling portion of the ICM.
This in turn allows us to construct a realistic scenario of
production, confinement, and transport of heavy elements,
from the scales of individual galaxies to the cluster-wide scale.

Spatial studies of nearby clusters with ASCA have also yielded a surprise; 
the CEE that was known previously in soft X-rays 
has also been detected in hard X-rays up to $\sim 10$ keV (subsection 3.1).
The revealed CEE properties in fact depend little on the X-ray energy.
We accordingly regard the CEE as a manifestation of particular shapes
of the gravitational potential at the cluster center (subsections 3.3 and 3.4),
rather than the ongoing ICM cooling.
This allows us for the first time to compare in close detail the X-ray 
determined potential shapes with those suggested by $N$-body simulations
(Navarro et al. 1996; Fukushige, Makino 1997; Moore et al. 1998; Ghigna et al. 2000),
or those derived from gravitational lensing measurements (e.g. Wu et al. 1998).

The present paper is meant to provide 
an overall summary of these new ASCA results,
mainly based on six PhD thesis works (Ikebe 1995; Matsushita 1997; 
Fukazawa 1997; Xu 1998;  Tamura 1998; Ezawa 1998).
Although many of these results have already been published individually,
here we assemble them together as building blocks,
and attempt to construct a novel comprehensive view 
of the physics in the cluster core regions. 
We devote Sections 2, 3, and 4 to the descriptions of
the ICM temperature structure, the spatial properties of the ICM,
and the ICM metallicity, respectively.
There, we incorporate brief comparison 
with the early Chandra and XMM-Newton results.
These results are combined together in Section 5.

In order to avoid substructures often seen in rich clusters, 
we mainly investigate relatively poor clusters,
with ICM temperature below $\sim 6$ keV. 
Unless otherwise stated, 
we use the $90$\% confidence error regions throughout,
and assume the cluster to be spherically symmetric.
For the sake of easy comparison with previous results,
we assume the Hubble constant to be $H_{0}=50~h_{50}$ km s$^{-1}$ Mpc$^{-1}$,
even though it somewhat differs from the most recent determinations. 
We also employ the solar abundance ratios from Anders, Grevesse (1989), 
with Fe/H$=4.68\times10^{-5}$ by number.
We denote the three-dimensional radius $R$,
while the projected radius $r$.

%===================== 2 ========================= 
\section{PROPERTIES OF THE CENTRAL COOL COMPONENT}
%===================== 2 ========================= 

The CCC properties were so far studied in a limited energy band below $\sim 3$ keV.
In addition, few such observations were performed with high spectroscopic resolution,
except those with the Einstein FPCS (e.g., Canizares et al. 1982),
the Einstein SSS (e.g., Rothenflug et al. 1984), 
and the BBXRT (e.g., MacKenzie et al. 1996).
These limitations, coupled with projection effects,
made it difficult to unambiguously quantify
how the ICM temperature decrease actually takes place 
(e.g., Thomas et al. 1987).
The superior energy resolution of ASCA,
together with its imaging capability up to $\sim 10$ keV,
have therefore enabled us to acquire novel information on the CCC phenomenon.

%----------- 2.1--------------------
\subsection{The Centaurus cluster}
%----------- 2.1 -------------------

The Centaurus cluster, 
with a relatively good circular symmetry, a low redshift (0.011),
and the most prominent CCC among nearby clusters 
(Matilsky et al. 1985; Allen, Fabian 1994; 
Fabian et al. 1994; Fukazawa et al. 1994; Ikebe 1995; Ikebe et al. 1999),
provides an ideal opportunity to investigate the ICM temperature structure.
According to the ROSAT data (Allen, Fabian 1994),
the CCC of this cluster is seen within a projected radius 
of $r \sim 5'$ ($\sim 100~h_{50}^{-1}$ kpc).
Fabian et al. (1994) described the ASCA results 
on the CCC in terms of the CF scenario.

The same ASCA data, taken with 
the GIS (Gas Imaging Spectrometer; Ohashi et al. 1996; Makishima et al. 1996) 
and the SIS (Solid-State Imaging Spectrometer; Burke et al. 1994; Yamashita et al. 1999),
were also analyzed by Fukazawa et al. (1994), Ikebe (1995), and Ikebe et al. (1999),
with increasing degree of sophistication.
Ikebe et al. (1999) also incorporated the ROSAT PSPC data.
These authors have established ``two-temperature (2T)'' picture 
of the ICM of this fascinating cluster:
the spectra accumulated over concentric annular regions around the 
cluster center can be fitted by a sum of cool and hot emission components,
with a temperature of $T_{\rm c} \sim 1$ keV and $T_{\rm h} \sim 4$ keV, respectively.
Since the relative contribution of the cool component clearly diminishes for $r > 7'$,
the cool component can be identified with the CCC.
In addition, these authors have obtained 
the following two important new findings.

One is that the spectra taken from the central ($r < 5'$) region contains 
not only the cool component, but also a significant amount of hot component,
which may have escaped detections by the low-energy imaging instruments.
Fukazawa et al. (1994) argued that the observed hot emission indeed comes 
mostly from the three-dimensional cluster core region rather than from 
the foreground or background off-center regions along the line of sight.
Therefore, the apparent 2T property does not result from projections of a 
single-phase ICM (i.e., the temperature being a single-valued function of $R$)
having a central temperature decrease,
but from multi-phase nature of the ICM in the core region of this cluster.
This view was reinforced by Ikebe (1995), who showed
that single-phase models with central temperature drops generally
underpredict the hard X-ray flux from the cluster center region.
Convincing support has been obtained by Ikebe et al. (1999);
the ICM within a three-dimensional radius
$R \sim 1.'5$ ($\sim 30~h_{50}^{-1}$ kpc) cannot be isothermal,
and at least two components of different temperatures must be involved there.

The other result concerns the number of different 
temperature components in the cluster core region.
The two temperatures from the 2T fit,  $T_{\rm c}$ and $T_{\rm h}$,
are both virtually constant as a function of $r$ (Fukazawa et al. 1994; Makishima 1994a),
which would not generally occur if many components with different 
temperatures and different angular distributions were involved.  
Furthermore, the ASCA and ROSAT spectra from the cluster center
have been fitted well ($\chi^2/\nu = 119/121)$ with the 2T model (Ikebe et al. 1999),
while the fit became unacceptable ($\chi^2/\nu = 159/122$)
when the cool component was replaced by the CF spectral model.
The conclusion remained unchanged even 
by applying a separate free excess absorption to the CF component, 
which gave only a slightly improved fit ($\chi^2/\nu =142/121$)
together with a best-fit excess column of $\sim 1.3 \times 10^{21}$ cm$^{-2}$.
Therefore, there seems to be only two major temperature components,
rather than a continuous distribution of emission measure 
as a function of temperature.

We hence regard the 2T modeling of Centaurus as physically meaningful,
rather than just a convenient description of the data.
Actually, the best-fit 2T solution obtained by Ikebe et al. (1999) can 
consistently and simultaneously describe the spatial and spectroscopic properties
of the existing ASCA GIS/SIS and ROSAT PSPC data of this cluster.
Specifically, it dictates the following features.
\begin{enumerate}
\item The two temperatures are $T_{\rm h}=3.9 \pm 0.1$ keV 
      and $T_{\rm c}=1.4 \pm 0.2$.
\item The cool phase, or CCC, is localized to within  
      $R \sim 5'$ ($\sim 100~h_{50}^{-1}$) of the cluster center, 
      and exhibits a 0.5--4 keV luminosity of 
      $1.0 \times 10^{43}~h_{50}^{-2}$ ergs s$^{-1}$. 
\item Assuming a pressure balance between the two phases,
      the CCC is calculated to have a mass of $(4-5) \times 10^{10}~ M_\odot$,
      and an average volume filling factor of 0.06--0.08 within $60~h_{50}^{-1}$ kpc.
      Thus, the central region is mainly occupied by the hot phase.
\item Toward the center, the ICM metallicity increases significantly 
      in both phases, up to $1-1.5$ solar abundances.
      The central excess iron contained in the two phases 
      amounts to $\sim 1 \times 10^9~ M_\odot$.
\item There is no evidence of excess X-ray absorption, 
      beyond typical upper limits of $\sim 1 \times 10^{21}$ cm$^{-2}$.
\item The gravitational potential becomes deeper at the center than a
      King-type potential, showing either a hierarchical structure
      or a central cusp.
\end{enumerate}
The first four items reconfirm or update the results 
obtained previously by Fukazawa et al. (1994) and Ikebe (1995),
while the second item agrees with Allen, Fabian (1994).

Thus, the first significant result of the present paper is
the successful review of the detailed 2T model for the Centaurus cluster,
with an implication that the cluster core volume is mostly filled 
with the hot phase even though there exists a significant CCC.

%%%%% Added %%%%%%%
The spatial co-existence of two distinct plasma components 
is also found in the case of diffuse Galactic X-ray emission, 
called Galactic ridge X-ray emission 
(GRXE; Koyama et al. 1986; Kaneda et al. 1997). 
Using the ASCA Galactic-plane survey data,
Kaneda (1997) analyzed the GRXE surface brightness variation 
along the Galactic longitude.
Employing cores-correlation and fractal dimension analyses,
he has unambiguously shown
that the 0.8--10 keV GRXE comprises {\it two} emission components, 
with characteristic temperatures of $\sim 0.9$  keV and 3--7 keV.

%-------------------- 2.2 -------------------------------------
\subsection{Two-temperature properties of a sample of clusters}
%-------------------- 2.2 -------------------------------------
%%%%% Added %%%
In clusters other than Centaurus, the CCC is generally not strong enough 
to unambiguously discriminate the 2T picture from other interpretations,
e.g., single-phased ICM with a temperature gradient,
or a highly multi-phased ICM with the temperature distributing over a wide range.
Nevertheless, in the case of the Centaurus cluster (subsection 2.1), 
the 2T model has been shown to be more appropriate than these alternatives.
Since there is no particular reason to regard Centaurus as exceptional,
we again adopt the 2T modeling as a working tool, and examine
whether it can describe the ASCA data of other clusters.
%%%%%%%%%%%%%%%%%%%%%%%%%%%

Actually, the ASCA SIS spectrum from a central region of 
the Perseus cluster, with a strong CCC,
was reproduced reasonably well with the 2T model,
although the ``hot-phase plus CF'' model was similarly successful (Fabian et al. 1994).
The 2T formalism has also been successful for 
central regions of the Hydra-A cluster and Abell~1795,
as reported by Ikebe et al. (1997b) and Xu et al. (1998) respectively.
The X-ray emission from M87, at the center of the Virgo cluster,
has also been modeled successfully by Matsumoto et al. (1996) in the same way,
using $T_{\rm h} = 2.4-3.2$ keV and $T_{\rm c} = 1.1-1.5 $ keV,
although the narrower separation between the two temperatures makes 
it rather difficult in this case to assess the uniqueness of the 2T formalism.
We here remember the conclusion derived previously by Canizares et al. (1982)
through an elaborate analysis of the Einstein FPCS data,
that the X-ray emitting plasma in M87 is highly multi-phased,
with the temperature distributed from $ \sim 0.3$ keV to $\sim 3$ keV or higher.
However, their conclusion is heavily dependent on the Fe-L complex,
of which the atomic model calculations later turned out to be problematic 
(e.g., Fabian et al. 1994; Matsushita et al. 1997).
Accordingly, reservations should be put on the interpretation by Canizares et al. (1982).

In addition to these individual attempts, 
Fukazawa (1997) and Fukazawa et al. (1998, 2000) took a statistical approach.
These authors used a sample of 40 clusters observed with ASCA,
and accumulated X-ray spectra for each object separately 
over the central (within $r \sim 100 ~h_{50}^{-1}$ kpc) and outer regions.
The sample definition and details of the data analysis 
are described in Fukazawa et al. (1998, 2000). 
For all clusters, the outer-region spectra were fitted successfully 
with a single-temperature plasma emission model.
In contrast, the same modeling failed for the
central-region spectra from about half of the objects in the sample.
These spectral fits have been improved significantly by employing the 2T formalism,
in which the value of $T_{\rm h}$ is set equal to the 
outer-region ICM temperature of the same cluster.
These clusters hence have statistically significant CCC,
for which the 2T formalism gives a successful description.
The 2T parameters derived by Fukazawa (1997) and Fukazawa et al. (2000)
are quoted in table~1, after discarding too rich ($T_{\rm h}>6.0$ keV)
or too poor ($T_{\rm h}<2.0$ keV) clusters:
the former criterion is to avoid objects with substructures.
Our reduced sample comprises 20 objects,
including, e.g.,  Centaurus, Virgo, Hydra-A, and Abell~1795,
but not Perseus which has $T_{\rm h} > 6$ keV.
Thus, the CCC temperature is found at $T_{\rm c} =1.1-2.2$ keV,
in agreement with the results on the Centaurus and Virgo clusters.
These results extend the validity of the 2T picture to a larger sample.

Also listed in table~1 is the ratio $Q_{\rm c}/Q_{\rm h}$,
where $Q_{\rm c}$ and $Q_{\rm h}$ are emission integrals 
of the cool and hot components, respectively, 
both integrated over the central $\sim 100 ~h_{50}^{-1}$ kpc.
These are related to the hot-phase and cool-phase luminosities,
$L_{\rm h}(\Delta E)$ and $L_{\rm h}(\Delta E)$ respectively,
emergent from that region in a specified energy band $\Delta E$, as
\begin{equation}
L_{\rm h}(\Delta E)= Q_{\rm h} \cdot \Lambda_{\rm h}(\Delta E) ~, ~~~
L_{\rm c}(\Delta E)= Q_{\rm c} \cdot \Lambda_{\rm c}(\Delta E) ~,
\label{eq:LhLc}
\end{equation}
where $\Lambda_{\rm h}(\Delta E) \equiv \Lambda(T_{\rm h},Z_{\rm h}; \Delta E)$ 
and $\Lambda_{\rm c}(\Delta E) \equiv \Lambda(T_{\rm c},Z_{\rm c}; \Delta E)$ 
are the band-limited cooling functions for the hot and cool phases, respectively,
with $Z_{\rm h}$ and $Z_{\rm c}$ denoting their metallicities.

Assuming the CCC to be confined within a volume $V$ 
with an average volume filling factor  $\eta$ (Fukazawa et al. 1994; Makishima 1994b),
the emission integrals can be written as
\begin{equation}
Q_{\rm h}= \left<n_{\rm h}^2 \right> (1-\eta) V~~, 
       ~~~Q_{\rm c}= \left< n_{\rm c}^2 \right> \eta
V~,  \label{eq:EMI}
\end{equation}
where $n_{\rm h}$ and $n_{\rm c}$ are 
plasma densities of the hot and cool phases respectively,
and the bracket means the spatial average over $V$.
Employing the assumption of pressure balance 
between the two phases (Fukazawa et al. 1994; Makishima 1994b) as
\begin{equation}
n_{\rm c} T_{\rm c} = n_{\rm h} T_{\rm h}~,
\label{eq:pbal}
\end{equation}
equation (\ref{eq:EMI}) yields
\begin{equation}
\eta = \left[ 1 + \left(\frac{T_{\rm h}}{T_{\rm c}}  \right)^2 
       \left(  \frac{Q_{\rm h}}{Q_{\rm c}} \right)   \right]^{-1}~
\label{eq:fillfact}
\end{equation}
[the same as eq.(9) of Ikebe et al. (1999)].

Taking the Centaurus cluster for example, the values listed in table~1,
$Q_{\rm c}/Q_{\rm h} \sim 1.0$ within $r=100~h_{50}^{-1}$ kpc 
and $T_{\rm h}/T_{\rm c} \sim 2.7$, 
give $\eta \sim 0.1$ via equation (\ref{eq:fillfact}).
This roughly agrees with the results from the more detailed analysis 
by Ikebe et al. (1999) quoted in subsection 2.1.
Similarly, in table~1, we give the value of $\eta$ for each cluster
calculated via equation (\ref{eq:fillfact}).
Thus, $\eta$ takes rather small values,
falling below 0.1 except for Abell~496, Centaurus, 3A~0335+096, and Virgo.
This implies that the core regions of clusters in table~1 
are occupied predominantly by the hot phase even if there exists a CCC.
Hereafter, we employ the 2T formalism as our standard tool,
and identify the CCC with the cool component.

From equations (\ref{eq:LhLc}), (\ref{eq:EMI}), and (\ref{eq:pbal}), 
we immediately obtain
\begin{equation}
\frac{L_{\rm c}(\Delta E)}{L_{\rm h}(\Delta E)}
  =  \left( \frac{Q_{\rm c}}{Q_{\rm h}} \right)
     \left( \frac{\Lambda_{\rm c}}{\Lambda_{\rm h}} \right)
  =  \left(\frac{T_{\rm h}}{T_{\rm c}} \right)^2
     \left(\frac{\Lambda_{\rm c}}{\Lambda_{\rm h}} \right)
     \frac{\eta}{1-\eta}~.
\label{eq:Lc/Lh}
\end{equation}
This allows us to relate $L_{\rm c}/L_{\rm h}$, 
$Q_{\rm c}/Q_{\rm h}$, and $\eta$ to one another.
Again taking Centaurus for example,
substitution of $\eta \sim 0.1$ into equation (\ref{eq:Lc/Lh})
yields $L_{\rm c}/L_{\rm h} \sim 0.8 \Lambda_{\rm c}/\Lambda_{\rm h}$,
and actual calculation of the cooling function
(using $T_{\rm h} =3.8$ keV and $T_{\rm c} =1.3$ keV,
and assuming $Z_{\rm h} \sim Z_{\rm c} \sim 1$)  
gives the $L_{\rm c}/L_{\rm h}$ ratio of 1.1, 0.16, and 0.68, 
in the 0.5--3, 3--10, and 0.5--10 keV bands, respectively;
these values are consistent with those calculated 
by Ikebe et al. (1999) for the Centaurus cluster.
Thus, in soft energies where $\Lambda_{\rm c} > \Lambda_{\rm h}$,
the CCC luminosity $L_{\rm c}$ of a cD cluster can be comparable
to the hot-phase luminosity  $L_{\rm h}$ from the cluster core region. 
In contrast, in higher energies or in a sufficiently wide energy range,
$L_{\rm c}$ obviously falls much below $L_{\rm h}$,
because $\eta <<1$ and $\Lambda_{\rm c} << \Lambda_{\rm h}$.

%%%% added %%%%
The Chandra and XMM-Newton observations are producing 
a series of interesting results on the ICM temperature structure,
that there is little X-ray emission component with the temperature
below a certain lower cutoff value that differs from object to object.
The cutoff temperature is reported to be
2.7 keV for Abell~1835 (Peterson et al. 2001),
1.5 keV for S\'ersic 159-03 (Kaastra et al. 2001),
and 2.5 keV for Abell~1795 (Fabian et al. 2001; Tamura et al. 2001)
which is close to our $T_{\rm c}$ (2.04 keV; table 1).
These results suggest 
that the X-ray spectra of each cluster are characterized by two temperatures,
i.e., the lower cutoff value and the outer-region averaged ICM temperature.
Thus, our 2T interpretation appears to be valid for these new data as well,
although it is yet to be examined
whether or not these new missions are detecting emission components 
with the temperature in between the two characteristic values. 
%%%%%%%%%%%%%%%%%%%%%%%%%%%%%%%

%---------------- 2.3 ----------------------------------
\subsection{The central cool component and cD galaxies}
%---------------- 2.3 ----------------------------------
Although found in many clusters,
the CCC is subject to a large scatter, from object to object, 
in its prominence relative to the whole cluster emission.
For example, compared to the Centaurus cluster,
the nearby cluster Abell~1060 is rather similar in distance, 
optical richness, ICM temperature, and the overall X-ray luminosity.
Nevertheless, its ICM is quite isothermal at $3.1 \pm 0.2$ keV right to the center, 
with little evidence for CCC.
More quantitatively, 
Tamura et al. (1996, 2000) and Tamura (1998) constrained
the 0.3--5 keV luminosity of CCC in this cluster
to be $< 6 \times 10^{41}~ h_{\rm 50}^{-2}$ ergs s$^{-1}$
over a central region of $r< 3'$ ($< 60~h_{\rm 50}^{-1}$ kpc).
This falls by more than an order of magnitude below that of the Centaurus cluster.
The same conclusion can be derived by comparing the two 
clusters in terms of the $Q_{\rm c} /Q_{\rm h} $ ratio given in table~1;
it is $< 0.1$ for Abell~1060, while $\sim 1$ for Centaurus.
In addition, Abell~1060 has long been know as a typical 
weak-CEE object (Jones, Forman 1984; see subsection 3.4).

What makes the two clusters so different?
Although CCC could be disrupted by mergers (e.g. Fabian 1994),
Abell~1060 is even more relaxed and rounded than Centaurus,
with least evidence for recent mergers.
Several authors, including Tamura et al. (1996), instead ascribe 
the difference to the properties of their central galaxies.
The Centaurus cluster is classified as Bautz-Morgan 
(B-M; Bautz, Morgan 1970; Abell et al. 1989) type I-II,
and hosts a single giant galaxy, NGC~4696, 
which completely dominates the central cluster region 
up to $\sim 200~h_{50}^{-1}$ kpc in radius.
In contrast, Abell~1060 has B-M type III,
and host two central elliptical galaxies, NGC~3309 and NGC~3311,
with a projected separation of only $\sim 35~h_{50}^{-1}$ kpc.
There is yet another bright spiral galaxy, NGC~3314, 
at a separation of $\sim 100~h_{50}^{-1}$ kpc from NGC~3311.
None of the three dominates the central region of Abell~1060.
Tamura et al. (1996) argue that these morphological properties are 
responsible for the X-ray difference between the two clusters;
the idea may be traced back to Jones, Forman (1984).

In order to investigate the above suggestion, 
we here subdivide the objects in table~1 into two subsamples,
cD and non-cD clusters.
In order to avoid tautology, we utilize solely the optical information:
we define cD clusters as those of B-M type I, I-II, or II,
while non-cD ones as those with B-M type II-III or III.
The only exception is the Virgo cluster; 
it is of B-M type III, but we classify it as a cD cluster,
because M87 is widely regarded as its cD galaxy.
Although the B-M type is unavailable for 3A~0335+096 and the Hydra-A cluster,
we classify them as cD clusters based on their Rood-Sastry classification 
as ``cD'' (Rood, Sastry 1971; Struble, Rood 1987).
Our classification is not much different from 
that of Fukazawa (1997) and Fukazawa et al. (2000),
who mainly utilized the Rood-Sastry morphology.

\begin{figure}[htb]
\vspace*{4cm}
\centerline{\psfig{file=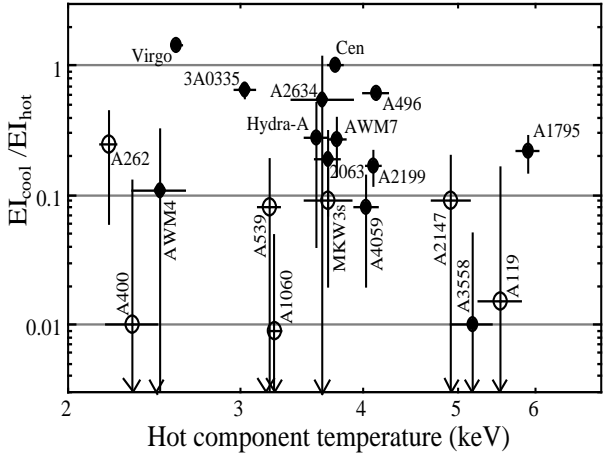,height=6cm,width=8cm}}
\caption{
Ratio of the emission integrals between the cool and hot ICM components,
calculated within $\sim 100\ h_{50}^{-1}$ of the cluster center
and plotted as a function of the hot-component temperature.
Filled and open symbols specify cD and non-cD clusters, respectively,
as defined in subsection 2.3.
Data refer to Table~1.
}
\label{fig:Qc/Qh}
\end{figure}

In figure \ref{fig:Qc/Qh}, we plot the $Q_{\rm c}/Q_{\rm h}$ ratio 
taken from table~1 as a function of $T_{\rm h}$.
It is thus clear that the cD clusters exhibit systematically
higher $Q_{\rm c}/Q_{\rm h}$ ratios than the non-cD ones.
We hence conclude that the CCC is selectively seen among cD clusters.
%which makes the third important result of the present paper.
Although the general association of CCC with cD galaxies has long been known 
(e.g., Jones, Forman 1984; Edge et al. 1992; Fabian 1994),
figure \ref{fig:Qc/Qh} allows us for the first time to quantitatively
discriminate cD and non-cD clusters in terms of the 2T picture.

%----------------------- 2.4 ----------------------------------
\subsection{Mass and luminosity of the central cool component}
%----------------------- 2.4 ----------------------------------

\begin{figure}[b]
\centerline{\psfig{file=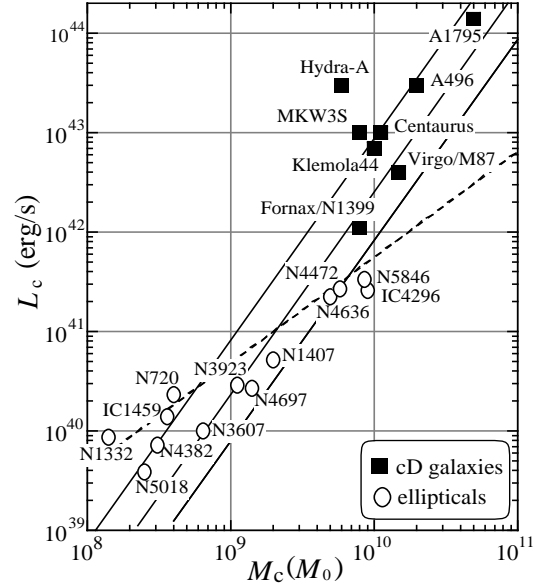,width=7cm,height=8cm}}
\caption{
The cool component mass $M_{\rm c}$ of cD clusters, 
plotted against the 0.5--3 keV cool component luminosity $L_{\rm c}$.
Data refer to Ikebe et al. (1999) for the Centaurus cluster,
Matsumoto et al. (1996) for Virgo/M87,
Xu et al. (1998) for Abell~1795,
Ikebe et al. (1997b) for the Hydra-A cluster,
Ikebe (1995) for Fornax/NGC~1399,
and Ikebe et al. (1997a) for Abell~496, Klemola 44, and MKW3s.
The values of $L_{\rm c}$ employed  here differ to some extent from those given in Table~1, 
mainly due to differences in the data integration radius.
The data for non-cD ellipticals (filled circles) are from Matsushita (1997).
The dashed line represents the scaling for a 1.0 keV plasma with a 0.5 solar abundance,
assumed to have a constant density of $2.5 \times 10^{-3}$ cm$^{-3}$
with variable occupation volumes. 
Solid lines indicate the case when the plasma has a uniform but variable density
and is confined within a constant spherical volume of radius 25 kpc, 
with a filling factor of 1 (bottom), 0.3 (middle), and 0.1 (top). 
}
\label{fig:McLc}
\end{figure}

The 2T modeling allows us to calculate the mass associated with the CCC 
(i.e., the cool-phase mass), $M_{\rm c}$,
as performed by Fukazawa et al. (1994) for Centaurus.
However for this purpose, 
we need to quantify the ICM density profile.
For this reason, we cannot derive $M_{\rm c}$ for all objects in table~1.
Instead, we refer to individual publications dealing with selected cD clusters, 
and compile the reported values of $M_{\rm c}$ 
in figure \ref{fig:McLc} against $L_{\rm c}$. 
Thus, the values of $M_{\rm c}$ of the selected cD clusters 
are at most several times $10^{10}~M_\odot$,
or only a few to ten percent of the stellar mass in a typical cD galaxy.

In figure \ref{fig:McLc}, 
we also show masses and X-ray luminosities of the 
ISM (inter-stellar medium; Forman et al. 1985; 
Trinchieri, Fabbiano 1985; Canizares et al. 1987)
associated with non-cD elliptical galaxies,
obtained by Matsushita (1997) using ASCA.
Thus, the CCC masses are only a few times higher 
than the ISM masses of the non-cD elliptical galaxies with 
the highest X-ray luminosities (e.g. NGC~4472, NGC~4636, NGC~5846, and IC~4296).
The CCC and the ISM emission are similar to each other 
in temperature and angular extent as well.
Furthermore, the CCC is predominantly seen around cD galaxies (subsection~2.2).
These comparisons reveal a close similarity between the CCC in cD clusters 
and the ISM emission from X-ray luminous elliptical galaxies
(see subsection 2.6 for a further discussion).
Nevertheless, in figure \ref{fig:McLc},
the CCC exhibits significantly higher luminosities than the ISM of ellipticals.

From equations (\ref{eq:LhLc}) and (\ref{eq:EMI}), 
the cool-phase luminosity scales as
\begin{equation}
L_{\rm c}(\Delta E) = \left<n_{\rm c}^2 \right> V \eta \cdot \Lambda_{\rm c}
                  ~~ \propto ~ M_{\rm c}^2 (V \eta)^{-1} \Lambda_{\rm c}
\label{eq:lumgeneral}
\end{equation}
where definitions of the symbols are the same 
as in equations (\ref{eq:LhLc}) and (\ref{eq:EMI}). 
Therefore, even $M_{\rm c}$, $V$, $T_{\rm c}$, and $Z$ 
(the latter two affecting $\Lambda_{\rm c}$) are kept constant, 
the luminosity can be increased by decreasing $\eta$.
The behavior of CCC in figure \ref{fig:McLc} can be explained in this way,
because the 2T interpretation implies 
that the CCC actually has very small values of $\eta$ (subsection 2.2).
Presumably, the CCC forms blobs or filaments around each cD galaxy,
compressed by the ambient hot ICM.

We thus conclude that the CCC of a cD galaxy has a close resemblance 
to the ISM of X-ray luminous elliptical galaxies 
with respect to the mass and spatial extent, 
but has a higher luminosity presumably 
because it is confined to achieve a lower filling factor.
The small values of $M_{\rm c}$ and the relatively high values $L_{\rm c}$
lead to rather short radiative cooling times for the cool component, as

\begin{footnotesize}
\begin{equation}
   \tau_{\rm c} = 1.7 \times 10^8 
                 \left(\frac{M_{\rm c}}{10^{11}\ M_\odot} \right) 
                 \left(\frac{L_{\rm c}}{10^{43} {\rm ergs \ s^{-1}} } \right)^{-1} 
                 \left(\frac{T_{\rm c}}{{\rm 1~keV}}  \right) ~~ {\rm yr}~~.
\label{eq:tauc}
\end{equation}
\end{footnotesize}

%----------- 2.5 -----------------------
\subsection{Amount of cool emission}
%----------- 2.5 -----------------------

The cool-component luminosities measured with ASCA,
summarized in table~1 and figure \ref{fig:McLc},
are considerably lower on average than was thought previously.
To compare our results with previous ones,
let us convert the bolometric cool-component luminosity 
$L_{\rm c}^{\rm bol}$ measured with ASCA
to the mass deposition rate $\dot{M}$ in terms of the CF scenario,
utilizing a theoretical relation of
\begin{equation}
L_{\rm c}^{\rm bol} = \frac{5 \dot{M} k T_{\rm h}}{2 \mu m_{\rm p}} ~~
\label{eq:CF_rates}
\end{equation}
(Fabian 1994)
where $k$ is the Boltzmann constant, $m_{\rm p}$ is the proton mass, 
and $\mu \sim 0.6$ is the mean molecular weight.
The ICM is assumed to start cooling from $T_{\rm h}$.

The Hydra-A cluster is one of the prototypical cooling-flow clusters
with the reported mass deposition rate reaching
$\dot{M} = 315^{+174}_{-82} ~ h_{50}^{-2}~ M_\odot$ yr$^{-1}$ 
(Edge et al. 1992), 
$ \sim 600~\ h_{50}^{-2}~ M_\odot$ yr$^{-1}$ (David et al. 1990),
or $ \sim 270~\ h_{50}^{-2}~ M_\odot$ yr$^{-1}$ 
(Allen, Fabian 1997; using ROSAT).
However, a joint 2T fit to the ASCA and ROSAT spectra
of the Hydra-A cluster yielded a rather weak cool component with 
$L_{\rm c} \sim 5 \times 10^{43} ~ \ h_{50}^{-2}$ erg cm$^{-2}$ s$^{-1}$ 
in the 0.5--3 keV band (Ikebe et al. 1997b).
Then, from equation (\ref{eq:CF_rates}) and $T_{\rm h} \sim 4$ keV,
and after bolometric correction, 
we obtain $\dot{M} \sim 60~\ h_{50}^{-2}~M_\odot$ yr$^{-1}$,
which falls by 5--10 times below the previous estimates.
%%% changed %%%
To examine if this result depends on our 2T modeling,
%%%%%%%%%%%%%%%%%%%%%%%%%%%%%%%%%%%
Ikebe et al. (1997b) further fitted the same spectra
directly with the CF spectral model of Mushotzky, Szymkowiak (1988),
and obtained a similarly low value of
$\dot{M}=  (60 \pm 30)~\ h_{50}^{-2}~M_\odot {\rm yr}^{-1}$.
Applying a separate excess absorption to the CF component 
did not change the result significantly.

A similar story applies to another well known cooling-flow cluster, Abell 1795.
With $T_{\rm h} = 6$ keV, this cluster is reported to have 
$\dot{M} =478 \ h_{50}^{-2} ~ M_{\odot}$ yr$^{-1}$ 
with EXOSAT (Edge et al. 1992),
or $\dot{M} \sim 500 \ h_{50}^{-2} ~ M_{\odot}$ yr$^{-1}$ 
with ROSAT (Briel, Henry 1996; Allen, Fabian 1997).
In contrast, the ASCA data yielded 
$L_{\rm c} = (1.4\pm 0.4) \times10^{44}~ h_{50}^{-2}$ erg s$^{-1}$  
in the 0.5--3 keV band (Xu et al. 1998), 
which translates to $\dot{M} \sim 150 \ h_{50}^{-2} ~ M_{\odot}$ yr$^{-1}$ 
via bolometric correction and equation (\ref{eq:CF_rates}).
Consistently, the CF-model fit to the ASCA SIS spectra yielded  
$\dot{M} \sim 131 \ h_{50}^{-2} ~ M_{\odot}$ yr$^{-1}$ (Fabian et al. 1994).
Thus, the values of $\dot{M}$ derived with ASCA 
are again $3 \sim 4$ times lower than the previous estimates. 

Yet a third interesting example is Abell~1060 discussed in subsection 2.3.
This cluster was thought to host a rather weak cooling flow with 
$\dot{M}  = (2.4-15) \ h_{50}^{-2} ~ M_{\odot}$ yr$^{-1}$ (Singh et al. 1988),
or $\dot{M}  = 6 \ h_{50}^{-2} ~ M_{\odot}$ yr$^{-1}$ (Stewart et al. 1984). 
However, the ASCA upper limit on the CCC (table~1) is so low
that it gives, after bolometric correction,
$\dot{M} < 1 \ h_{50}^{-2}~ {\rm yr}^{-1}$ 
in terms of equation (\ref{eq:CF_rates}).
Again, this falls by an order of magnitude below the previous estimates.

\begin{figure}[t]
\centerline{\psfig{file=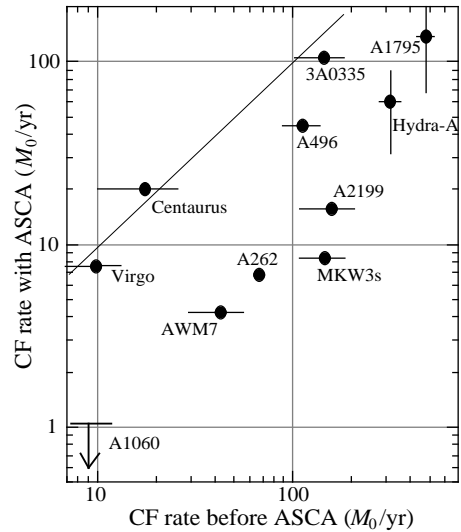,width=6cm,height=7cm}}
\caption{Comparison between cooling flow rates measured with ASCA,
against those reported previously (Edge et al. 1992).
The ASCA values of $\dot{M}$ were taken from
Ikebe et al. (1997b) for Hydra-A, 
Ikebe et al. (1999) for Centaurus, 
and Xu et al. (1998) for Abell~1795;
other ASCA  values were calculated by substituting $L_{\rm c}$ in Table~1,
after bolometric correction, into equation (8).
Although ASCA errors are shown only for Abell~1795 and the Hydra-A cluster,
those for the remaining objects are similar.
}
\label{fig:CFrates}
\end{figure}

Including these particular examples,
we compare in figure \ref{fig:CFrates} the mass deposition rates 
determined spectroscopically with ASCA,
against those derived in soft X-rays
mainly employing the surface brightness profiles.
Although the two estimates agree on the clusters with the strongest CCC
(Centaurus, Virgo, and 3A~0335+096),
for other clusters the ASCA values fall systematically and significantly below the previous ones.
This indicates that the past imaging spectroscopy in soft X-rays 
(especially, the commonly used deprojection analysis)
has considerably overestimated the cooling mass deposition rates of clusters.
%This is the fifth new result constituting the present paper.

%%%%%% Added and shifted %%%%%
This conclusion is being reconfirmed by a series of new results from Chandra and XMM-Newton.
For example, analysis of the Chandra data of the Hydra-A cluster
has led McNamara et al. (2000) to derive 
$\dot{M}=  (34 \pm 5)\ ~M_\odot {\rm yr}^{-1}$ within $R=74$ kpc,
for $H_0 = 70$ km s$^{-1}$ Mpc$^{-1}$:
the value would not greatly exceed $100~M_\odot {\rm yr}^{-1}$ 
even if we integrate up to $\sim 200$ kpc.
The spectra do not require excess absorption 
above the Galactic column of $2 \times 10^{20}$ cm$^{-2}$.
These results have further been detailed by David et al. (2001).
Similar deficits of the cool plasma are reported on Abell~1795 (Tamura et al. 2001),
Abell~1835 (Peterson et al. 2001), and S\'ersic 159-03 (Kaastra et al. 2001),
all being typical CF clusters.
The reported scarcity of low-temperature component is
independent of the validity of our 2T picture.
%%%%%%%%%%%%%%%%%%%%%%%%

%--------------------- 2.6 ---------------------------
\subsection{The nature of the central cool component}
%--------------------- 2.6 ---------------------------
%%%%%%% shortened %%%%
In the high-quality ASCA spectra 
taken from central regions of cD clusters,
we have thus successfully resolved the CCC as a spectroscopic cool component.
The CCC, or the cool phase, is localized around the cD galaxy (subsection 2.3),
presumably forming blobs or filaments immersed in the vast sea 
of the hot phase (subsection 2.4).
%%%%%%%%%%%%%%%%%%%%%%%%
This picture apparently agrees with the scenario 
of inhomogeneous CF (Nulsen 1986; Fabian 1994).
However, invoking the CF hypothesis in its original form may no longer be appropriate,
because we have at the same time discovered 
that the ICM deposition rate due to radiative cooling was significantly
overestimated in the past, by up to an order of magnitude (subsection 2.5).
Then, what is the nature of CCC?

Considering that non-cD elliptical galaxies in clusters 
(e.g., NGC~4406 and NGC~4472 in the Virgo cluster; 
Forman et al. 1985; Awaki et al. 1994) usually possess their own ISM, 
a cD galaxy, which is far less subject to ram-pressure stripping, 
must have its own ISM as well.
Then, by extending the argument of subsection 2.4, 
it is most natural to interpret the central cool phase as the ISM associated 
with the cD galaxy (Makishima 1996, 1997b, 1999b; Ikebe et al. 1999),  
rather than as a cooling portion of the ICM.
Specifically, we have the following pieces of evidence in support of this view; 
(1) the CCC is statistically associated with cD galaxies (subsection 2.3, figure \ref{fig:Qc/Qh});
(2) the CCC is positionally centered on cD galaxies (e.g., Lazzati, Chincarini 1998)
even if they are sometimes offset from the dynamical cluster centers;
(3) both the CCC and the ISM emission have a similar angular extent of several tens kpc;
(4) the CCC temperature, $T_{\rm c} =1\sim 2$ keV, 
is close to that of an elliptical's ISM, which in turn 
is consistent with the potential depth of giant elliptical galaxies; 
(5) the estimated mass of the cool phase is comparable to the ISM mass 
of ordinary elliptical galaxies (subsection 2.4, figure \ref{fig:McLc}).  
The only difference is the considerably higher X-ray luminosity of CCC 
compared to the ordinary ISM emission, 
but this can be explained as a result of compression by the hot phase (subsection 2.4).
%%%%%%%% changed %%%
The important point is that the CCC has a lower temperature 
because it reflects the shallower potential depth of a cD galaxy,
rather than because it is radiatively cooling.
%%%%%%%%%%%%%%%%%%%%%%%%%%%%%%%%%%%%%%%%%%%%%%%%%%%%%%%%%
Later, we reinforce this interpretation from metallicity arguments.

The above new picture, however, involves two immediate problems.
One is theoretical;
what sort of energy input (heating) 
prevents the cool phase from radiative cooling collapse,
and how are the two phases kept thermally insulated from each other?
The heating mechanism to be invoked,
though not required to have as large a luminosity as thought previously,
must be rather efficient to balance 
the rapid radiative cooling of equation (\ref{eq:tauc}).
Furthermore, the heating must balance the cooling in a stable manner,
in order to sustain the cool phase.
These are indeed the fundamental issues which underlie the CF hypothesis (Fabian 1994);
we briefly discuss them in Section 5.

The other problem is observational;
what caused the previous overestimates of $\dot{M}$?
An obvious possibility is ignorance of the dominance 
of the hot phase in the cluster core regions,
which in turn was due to the limited energy range and insufficient 
spectral resolution of the previous imaging X-ray observations.
Observers tend to attribute all the X-ray flux measured from the core region,
which is in fact dominated by the hot component, to the cooling flux.
However, this idea alone is inadequate to fully answer the question.
Actually, even using the ROSAT PSPC data alone,
discrepant values of $\dot{M}$ are sometimes assigned 
to a single object depending on the data analysis method.
For example, $\dot{M} < 80~M_\odot$ yr$^{-1}$ is obtained 
through analysis of the PSPC spectra of Abell~4059,
whereas the surface brightness analysis of the same PSPC data yields 
$\dot{M} \sim 184~M_\odot$ yr$^{-1}$ for the same object (Huang, Sarazin 1998).
Similarly, values of $\dot{M}$ as high as $\sim 10^3~M_\odot$ yr$^{-1}$ are
reported for some distant clusters via deprojection analysis of the ROSAT data,
even though the spectral evidence for CCC is poor (e.g., Schindler et al. 1997).
Therefore, the remaining clue may reside in the surface brightness profiles,
especially the CEE phenomenon.
This urges us to study, in the next Section,
spatial properties of the cluster X-ray emission using ASCA.

%====================== 3 ================================
\section{GRAVITATIONAL POTENTIAL SHAPE IN CLUSTER CENTERS}
%====================== 3 ================================

Imaging soft X-ray observations have established the CEE 
(central excess emission) as a ubiquitous phenomenon among cD clusters.
For example, analysis of the ROSAT X-ray surface 
brightness profiles for a large sample of clusters, 
using single-$\beta$ modeling, reveals a tight correlation between 
$\beta$ and the core radii (Pownall, Stewart 1996; Neumann, Arnaud 1999). 
This indicates the prevalence of CEE,
because both these quantities are known to take rather small values
when a single-$\beta$ model is forced to fit a
brightness profile with a significant CEE (Makishima 1995).
However, before ASCA, there has been essentially
no CEE investigation at energies above $\sim 3$ keV.

%----------------- 3.1 -----------------------------
\subsection{Central excess emission in hard X-rays}
%----------------- 3.1 -----------------------------

Generally, the X-ray volume emissivity $\epsilon$ of the ICM 
in a given energy band $\Delta E$ is expressed as 
\begin{equation} 
 \epsilon(R;\Delta E) = n(R)^2 \cdot \Lambda(T,Z;\Delta E) ~.
\label{eq:emissivity}
\end{equation}
For reference, volume integration of this equation for each ICM phase,
taking into account the filling factor,
together with equation (\ref{eq:EMI}), yields equation (\ref{eq:LhLc}).
Thus, $\epsilon$ is directly proportional to $n^2$,
and hence the CEE phenomenon implies an excess ICM density 
in the cluster core region above the prediction of a $\beta$ model.
This statement remains valid even if the ICM is deviated significantly from isothermality,
since the cooling function $\Lambda(T,Z;\Delta E)$ depends 
only weakly on $T$ in the relevant parameter regime.

The implied excess ICM density in the cluster core region, in turn, 
may be produced by either of the following two mechanisms.
One is that the ICM pressure distribution has a flat core without central excess,
and the increase in $n(R)$ is compensated by a decrease in the ICM temperature,
such as is invoked in the CF hypothesis.
The other possibility is that the ICM is relatively isothermal,
and hence the ICM pressure itself exhibits an excess in the central region.
So far, the attention has been paid predominantly to the former mechanism,
based on a belief that the CEE phenomenon is 
a spatial counterpart to the CCC phenomenon.
Observationally, in energies below $\sim 3$ keV,
these alternatives produce very similar effects and are difficult to distinguish.
However, in sufficiently high energies, their effects become distinct:
the former predicts a central deficit in the hard X-ray brightness
because the hot phase is displaced by the increasing cool-phase contribution,
while the latter obviously predicts a central excess in hard X-rays.

In order to examine the issue,
Ikebe et al. (1997b) and Xu et al. (1998) respectively analyzed 
the ASCA data of the Hydra-A cluster and Abell~1795, 
both known to host strong soft-band CEE 
(e.g. Jones, Forman 1984; David et al. 1990).
By fully taking into account the instrumental responses of ASCA,
they have discovered that these clusters surprisingly exhibit,
even in high energies above $\sim 3$ keV,
a clear CEE above the $\beta$ models describing their outer-region X-ray profiles.
Furthermore, in both objects, 
the relative prominence of CEE has been found to be roughly energy independent,
from the ROSAT range (0.2--2 keV) up to $\sim 10$ keV
(figure~9 of Ikebe et al. 1997b and table~3 of Xu et al. 1998).
According to Xu (1998), 
the same is approximately true of AWM7 and Abell~2199,
which are also known to show prominent CEE in soft X-rays.
Since the CCC contribution in these objects is negligible above $\sim 3$ keV,
the approximately color-independent CEE unambiguously indicates 
the presence of central excess pressure,
supporting the idea
that the ICM pressure exhibits an increase at small radii 
rather than that the ISM pressure has a flat core
(and hence the gas temperature decreases).

Then, what happens in clusters with very strong CCC, such as Centaurus,
in which the soft-band CEE is also strong (Matilsky et al. 1985)?
Using ASCA, Ikebe et al. (1999) have revealed
that the CEE of Centaurus actually gets weaker as the energy increases,
but it does not vanish even in the highest ASCA band;
the relative prominence of the central excess is roughly the same
between the 4.5--6.1 keV and 7.1--10 keV bands (figure~4 of Ikebe et al. 1999).
Therefore, there must exist central excess pressure,
which is strong enough to cause the hard-band CEE
against the displacement by the CCC.
The stronger CEE in lower energies ($< 4$ keV) can be 
attributed to the additional contribution from the CCC.
Xu (1998) has derived a very similar results  from 3A~0335+096,
which also host a strong CCC.

Thus, from more than half a dozen cD clusters with soft X-ray CEE,
we have detected the CEE also in hard X-rays with energies above 3 keV.
Since they are typical CEE objects,
we can generally ascribe the CEE phenomenon primarily to a central excess pressure. 
The central ICM temperature decrease (or the CCC) is concluded
to contribute partially to the CEE only in soft X-rays,
and only in limited objects such as Centaurus and 3A~0335+096.
In terms of the 2T picture developed in Section 2,
we can rephrase that the {\em hot-phase emissivity profile} of these cD clusters
exhibit a central excess above the $\beta$-model distribution.
These novel findings drastically renew our understanding of the CEE phenomenon.
For example, the strong CEE observed from the core regions
of several distant clusters (e.g. Schindler et al. 1997),
which is usually attributed to CFs,
may in reality be the potential shape effect,
since these objects do not necessarily exhibit strong CCC.

\begin{figure}[htb]
\centerline{\psfig{file=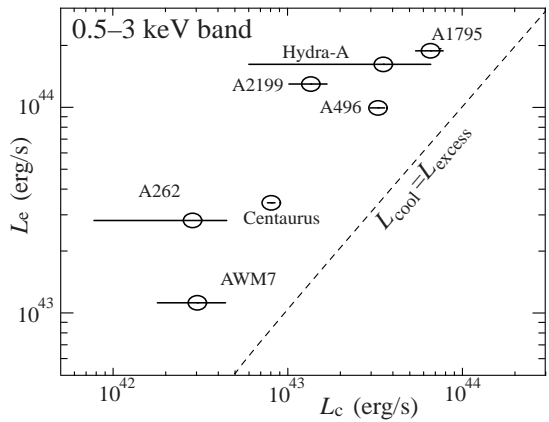,width=7.5cm}}
\caption{
The 0.3--5 keV cool-component luminosity $L_{\rm c}$ 
for a sample of cD clusters measured with ASCA (taken from Table~1),  
shown against their central excess luminosity $L_{\rm e}$.
The value of $L_{\rm e}$ refers to the integrated 0.5--3 keV luminosity
of the narrower $\beta$-component found by Mohr et al. (1999)
through their double-$\beta$ fitting to the ROSAT PSPC profile;
the integration is performed up to the maximum radius where the data exist,
but the results are not very sensitive to the upper boundary.
}
\label{fig:LcLe}
\end{figure}

Figure \ref{fig:LcLe} compares 
the CCC luminosity $L_{\rm c}$ of representative objects 
in the 0.5--3 keV band measured with ASCA (taken from table~1),
against the CEE luminosity $L_{\rm e}$ calculated in the same band
based on the ROSAT data analysis by Mohr et al. (1999).
Thus, $L_{\rm e}$ considerably exceeds $L_{\rm c}$,
even though the objects plotted here are known to host a rather strong CCC.
This result can be explained in the following way.
Generally, we may write the overall luminosity from the cluster center region 
in a given energy band $\Delta E$ as
$L_{\rm h}^{(0)}(\Delta E)+L_{\rm h}^{(1)}(\Delta E)+L_{\rm c}(\Delta E)$,
where $L_{\rm h}^{(0)}(\Delta E)$ refers to the cluster-wide hot emission
of which the radial profile is expressed with a $\beta$ model,
while $L_{\rm h}^{(1)}(\Delta E)$ describes the central excess in the hot phase.
In this three-term expression, the sum of the first two terms gives 
the hot-phase luminosity $L_{\rm h}(\Delta E)$ from the cluster center region,
while the sum of the last two terms yields $L_{\rm e}(\Delta E)$. 
Then, the implication of figure \ref{fig:LcLe} is 
that $L_{\rm h}^{(1)}(\Delta E)$ of these objects 
considerably exceeds $L_{\rm c}(\Delta E)$ even in the soft X-ray band.
This reinforces our previous inference made in subsection 2.2,
that the cluster core volume is dominated by the hot phase.
In the hard X-ray band (e.g., 3--10 keV),
we obviously expect $L_{\rm c}$ to become negligible compared to $L_{\rm e}$.

%------------------------ 3.2 -------------------------------------
\subsection{Excess central mass as the origin of CEE}
%------------------------ 3.2 -------------------------------------

To explore the implication of the central excess pressure, 
let us remember that the pressure $p(R)$ of hydrostatic ICM 
confined by gravity satisfies the relation (Sarazin 1988)
\begin{equation} %eq.2
M(R) = - \frac{kTR}{G \mu m_{\rm p}} 
  \left( \frac{{\rm d} \ln p}{{\rm d} \ln R} \right)~,
\label{eq:M(R)}
\end{equation}
where $M(R)$ is the total gravitating mass within $R$,
$G$ is the gravitational constant, 
and definitions of $k$, $m_{\rm p}$, and $\mu$ 
are the same as in equation (\ref{eq:CF_rates}). 
When the pressure profile exhibits a central excess
and the ICM is approximately isothermal, we may write
\begin{equation} %eq.3
p(R) = p_0(R) \xi(R) ~,
\label{eq:excess_p}
\end{equation}
where $p_0(R)$ is the pressure distribution described with a $\beta$ model
(or its equivalent, having a flat core),
and $\xi(R)>1$ is a non-dimensional factor describing the CEE
which tends to unity for large values of $R$.
Substituting equation (\ref{eq:excess_p}) into equation (\ref{eq:M(R)}), we obtain
\begin{equation} %eq.4
M(R) = M_0(R) + \delta M(R)~.
\label{eq:M+dM}
\end{equation}
Here $M_0 =  - (kTR / G \mu m_{\rm p})( {\rm d} \ln p_0 / {\rm d} \ln R )$
denotes the ``unperturbed'' total gravitating mass distribution
expressed with a King-type solution having a flat core,
while 
\begin{equation} 
\delta M(R) = - \frac{kTR}{G \mu m_{\rm p}} 
         \left( \frac{{\rm d} \ln \xi}{{\rm d} \ln R} \right)
\label{eq:dM}
\end{equation}
is an excess mass associated with the central excess pressure.

Numerically, 
the ROSAT (Allen, Fabian 1994) and ASCA (Ikebe et al. 1999) results 
both indicate an approximate radial dependence of $n(R) \propto R^{-1}$,
over the $R=(10-100)~h_{50}^{-1}$ kpc region of the Centaurus cluster.
This in turn implies $p(R) \propto R^{-1}$ 
considering the dominance of the hot component.
In contrast, the outer-region potential of Centaurus 
can be described by a King-type profile with a
core radius of $R_{\rm c} \sim 7' = 140~h_{50}^{-1}$ kpc
(Matilsky et al. 1985; Ikebe et al. 1999),
inside which $p(R)$ would be constant if there were no CEE.
From these facts we may approximate as $\xi(R) \propto R^{-1}$ 
in equation (\ref{eq:dM}), 
to obtain $-{\rm d} \ln \xi(R) / {\rm d} \ln R \sim 1$.
Taking this as a representative case, we can numerically write as
\begin{equation} 
\delta M  = 1.5 \times 10^{13} \left(\frac{T_{\rm h}}{4~ {\rm keV}} \right)
               \left(\frac{R}{100~{\rm kpc}} \right)~ M_\odot~~~.
\label{eq:dMnum}
\end{equation}
A more exact treatment considering the cool phase is found in Ikebe et al. (1999).

We can thus infer that a cluster with a hard X-ray CEE hosts an excess 
mass $\delta M$ of order $10^{13}~M_\odot$ at its center, 
above the King-type mass distribution. 
This $\delta M$ produces an additional potential drop at the center, 
which is superposed upon a flat core of the King-type potential profile.
The potential drop in turn attracts an excess amount of hot ICM,
thus producing the central excess pressure,
and hence the hard X-ray CEE
(Makishima 1995, 1998; Ikebe et al. 1996, 1997b; Xu et al. 1998).

The detections of CEE in hard X-rays are a second reason
why the ICM mass deposition rate was previous overestimated (subsection 2.5).
That is, the deprojection analysis usually proceeds by 
calculating the ICM density profile directly from the observed 
X-ray surface brightness profiles via equation (\ref{eq:emissivity}), 
and when the spectral information is inadequate,
the temperature profile is derived assuming an  
ICM pressure equilibrium in a plausible gravitating mass distribution,
by using equation (\ref{eq:M(R)}).
Therefore, if the central excess mass (or potential drop)
is not properly taken into account or underestimated,
the CEE immediately leads to a false central temperature decrease,
and hence to an overestimate of $\dot{M}$.

%---------------------- 3.3 ---------------------------
\subsection{Hierarchical gravitational potential}
%---------------------- 3.3 ---------------------------

The excess mass $\delta M$ thus found at the center of a cD cluster
can most naturally be attributed to the total gravitating mass 
associated with the cD galaxy (Makishima 1998),
because the CEE is nearly always centered on the 
position of the cD galaxy (Lazzati, Chincarini 1998)
even if it is somewhat offset from the dynamical cluster center,
and because the excess mass described with equation (\ref{eq:dMnum}) 
is reasonable as the total mass of a giant elliptical galaxy.
In other words, the gravitational potential exhibits a {\it hierarchy}, 
between a wider component associated with the entire cluster,
and a narrower one associated with the central galaxy. 
Such a picture has already been considered by various authors
(e.g., Thomas et al. 1987) in their deprojection analysis,
but has not been recognized generally as the origin of the CEE.

The cluster/galaxy hierarchy in the total mass distribution has first been 
recognized clearly in ASCA data of the Fornax cluster and its cD galaxy NGC~1399 
by Makishima (1995, 1996), Ikebe (1995), and Ikebe et al. (1996).
These authors successfully reproduced the radial X-ray  
brightness profile observed with ASCA, 
employing a ``double-$\beta$'' emissivity distribution of the form 
\begin{equation} %eq.6
\epsilon(R; \Delta E) = \epsilon_0(R; \Delta E) + \epsilon_1(R; \Delta E) ~~,
\label{eq:double_beta}
\end{equation}
where $\epsilon_0$ and $\epsilon_1$, both employing three-dimensional $\beta$ models,
describe the cluster component and the cD galaxy component, respectively.
This decomposition is essentially the same as what we have done in subsection 3.1 
($L_{\rm h} = L_{\rm h}^{(0)}+L_{\rm h}^{(1)}$).

In practice, we fit the observed radial X-ray surface brightness profile
by a sum of two 2-dimensional $\beta$-model components,
which are obtained by analytically projecting 
equation (\ref{eq:double_beta}) onto two dimension.
This fixes $\epsilon_0(R)$ and $\epsilon_1(R)$,
because a two-dimensional $\beta$ model is uniquely 
related to its three-dimensional counterpart.
Once the emissivity profile of equation (\ref{eq:double_beta}) is thus determined,
we can convert it into density profile via equation (\ref{eq:emissivity}),  
and further into pressure profile assuming approximate isothermality.
Obviously, the factor $\xi$ in equation (\ref{eq:excess_p}) can be given
in this case as $\xi(R) = \sqrt{1 + \epsilon_1(R)/\epsilon_0(R)}$,
and equation (\ref{eq:dM}) shows 
that $\epsilon_1$ is responsible for the central excess mass.
From the derived pressure profile,
we can calculate the gravitating mass profile $M(R)$ via equation (\ref{eq:M(R)}).

\begin{figure}[htb]
\centerline{\psfig{file=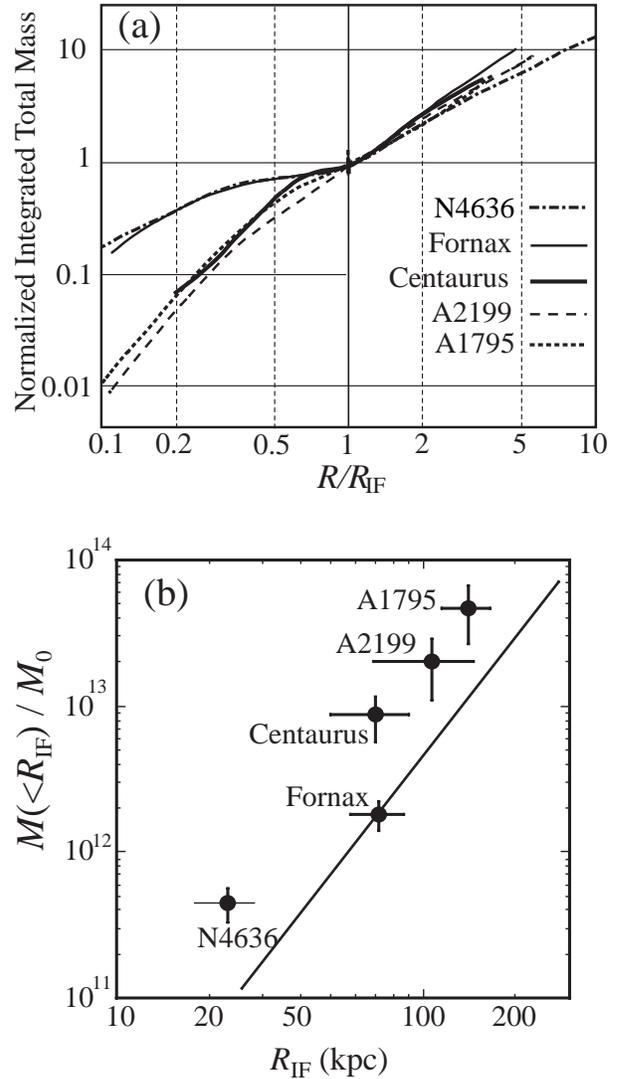,width=8cm}}
\caption{
Summary of the cases in which the hierarchical gravitational potential 
has been observed;
the Fornax cluster (for a distance of 20 Mpc; Ikebe et al. 1996),
NGC~4636 (for a distance of 17 Mpc; Matsushita et al. 1998),
the Centaurus cluster (Ikebe et al. 1999), 
Abell~1795 (Xu et al. 1998), and Abell~2199 (Xu 1998).
For the latter three objects, $H_0 = 75$ km s$^{-1}$ Mpc$^{-1}$ is assumed.
(a) The integrated curves of the total gravitating mass.
Abscissa, the 3-dimensional radius, 
is normalized to the interface radius $R_{\rm IF}$ 
(see text subsection~3.3 for definition), 
while ordinate is normalized to the mass contained within $R_{\rm IF}$.
Errors are not shown for clarity.
(b) Relation between $R_{\rm IF}$ and the total gravitating mass contained within it.
These quantities are used to normalize the mass curves in panel (a).
The straight line indicates a prediction for the universal halo 
(see text for detail).}
\label{fig:M(r)}
\end{figure}

The mass profile of the Fornax cluster, thus calculated by Ikebe et al. (1996),
is reproduced in figure \ref{fig:M(r)}a.
There, the 3-dimensional radius is normalized to ``interface radius'',
$R_{\rm IF} = 72 \pm 14$ kpc,
which is defined as the cross-over point
of the two emissivity terms in equation (\ref{eq:double_beta}).
The mass curve reveals a  ``shoulder'' like structure at $R \sim R_{\rm IF}$.
Outside $R_{\rm IF}$, we observe the cluster-scale mass distribution,
corresponding to the first term of equation (\ref{eq:double_beta}).
Inside $R \sim R_{\rm IF}$, in contrast, 
there appears an additional mass component 
due to the second term of equation (\ref{eq:double_beta}),
which can be identified with the excess mass associated with the cD galaxy.
Thus, $R_{\rm IF}$ may be regarded as an interface between 
territories of the cD galaxy and the overall cluster (Makishima 1996).
Substitution of $R=R_{\rm IF} \sim 72$ kpc and $T = 1.1$ keV
into equation (\ref{eq:dMnum}) yields $\delta M \sim 3 \times 10^{12}~M_\odot$,
in rough agreement with the data point in figure \ref{fig:M(r)}b.

The same double-$\beta$ modeling has been applied successfully to the 
ASCA GIS radial profiles of the Hydra-A cluster (Ikebe et al. 1997b), 
Abell~1795 (Xu et al. 1998), Abell~2199, AWM7, and 3A~0335+096 (Xu 1998),
all exhibiting the clear hard X-ray CEE (subsection 3.1).
For several of these objects, we show in figure \ref{fig:M(r)}a 
the rescaled mass curves derived in the same way as for the Fornax cluster.
The results on the Centaurus cluster (Ikebe et al. 1999) properly 
takes into account the effect of CCC by modifying equation (\ref{eq:double_beta}).
Thus, the mass curves again bear the characteristic feature at $R \sim R_{\rm IF}$.
Although in some cases (e.g., the Centaurus cluster; see subsection 3.5)
the apparent slope change in the mass curve at $R \sim R_{\rm IF}$ might be 
an artifact caused by the particular form of equation (\ref{eq:double_beta}),
the feature is real in Abell~1795 (Xu et al. 1998) as well as in Fornax.
Therefore, the hierarchical potential shape is 
suggested to be ubiquitous among these objects.

The potential hierarchy is observed from poorer systems as well. 
The best example is the giant elliptical galaxy NGC~4636,
around which a large-scale (up to $\sim 300$ kpc) ambient 
X-ray emission was detected with ROSAT (Trinchieri et al. 1994).
Using ASCA, Matsushita (1997) and Matsushita et al. (1998) have shown
that the large-scale emission in fact comes from hot gas 
trapped in the large-scale potential of an ``optically dark group'',
for which NGC~4636 plays the role of ``mini-cD'' galaxy.
The X-ray surface brightness was described 
successfully by equation (\ref{eq:double_beta}), 
and the calculated mass curve in figure \ref{fig:M(r)}a 
again bears the same feature at $R_{\rm IF} \sim 23$ kpc,
%%% modified %%%%%%%
which is not an artifact caused by the particular model form
of equation (\ref{eq:double_beta}) (Matsushita et al. 1998).
There is evidence (Matsushita 1997; Matsushita et al. 2000)
that X-ray luminous elliptical galaxies can generally be regarded 
as mini-cD galaxies of some larger-scale potential structures,
whereas X-ray dim ellipticals lack such outer potential envelopes.
%%%%%%%%%%%
The ``X-ray overluminous elliptical galaxies'' (Vikhlinin et al. 1999) 
may be similar in concept to our ``mini-cD'' galaxies. 

Figure \ref{fig:M(r)}b summarizes the values of $R_{\rm IF}$
and the total gravitating mass within it,
which are used to normalize the mass curves in figure \ref{fig:M(r)}a.
The results on Fornax (Ikebe et al. 1996) and NGC~4636 (Matsushita et al. 1998)
assume the distances of 20 Mpc and 17 Mpc, respectively, 
which are consistent with the Hubble constant of 
75 km s$^{-1}$ Mpc$^{-1}$ ($h_{50} = 1.5$).
For consistency, in figure \ref{fig:M(r)} we therefore employ  
$h_{50} = 1.5$ for the remaining three more distant objects,
unlike elsewhere in this paper.
We see a tight positive correlation between $R_{\rm IF}$ and the enclosed mass.

The nested two-component emissivity profiles have been found even in the ROSAT data.
Mulchaey, Zabludoff (1998) report that such an X-ray  
emissivity profile is common among galaxy groups.
Mohr et al. (1999) successfully employed the double-$\beta$ modeling
to express radial X-ray profiles of a large number of clusters observed with ROSAT.
%%%% Modified %%%%%%%%%%%%%%
Furthermore, the latest Chandra data of the Hydra-A cluster clearly reveal 
a hierarchical surface brightness distribution with $R_{\rm IF} \sim 100$ kpc
(David et al. 2001).
These reports suggest that the hierarchical potential shape is 
ubiquitous among self-gravitating systems,
%%%%%%%%%%%%%%%%%
although the effects found in the ROSAT data
may be partially attributable to the ICM temperature gradient.

%%%% Modified %%%%%%%%%%%%%%
We can strengthen our discovery of the hierarchical 
potential structure around cD galaxies 
by referring to several optical results suggestive of the same effect.
One is the large extended stellar envelope of cD galaxies,
of which the surface brightness profile deviates significantly from the 
standard de Vaucouleurs law (e.g. Schombert 1986; Johnstone et al. 1991);
the stellar envelope suggests the presence of a larger-scale 
potential structure surrounding the cD's own potential.
%%%%%%%%%%%%%%%%%
%%%% Added %%%%%%%%%%%%%%
The other is the outward increase in the stellar velocity dispersion of a cD galaxy,
which led Dressler (1979) to model the gravitational potential of a cD cluster
as a sum of hierarchically-nested {\em three} King models.
Our hierarchical potential structure is essentially identical to the Dressler's modeling,
although we need only two spatial components instead of three.
%%%%%%%%%%%%%%%%%

As shown so far, the CEE often (if not always) results from 
a nested concentric hierarchy in the gravitational potential,
formed by the cluster and its cD galaxy.
We further discuss its implication in subsection 3.5.
%% The sentences below were moved to 3.5.

%----------------------- 3.4 ----------------------------
\subsection{Central cusp in the gravitational potential}
%----------------------- 3.4 ----------------------------

The CEE is known to be very weak or nearly absent 
in some clusters (Jones, Forman 1984), 
e.g. the Coma cluster, Abell~400, and Abell~1060,
which mostly lack cD galaxies.
Then, what is the gravitating mass distribution in these non-cD clusters?
Does a King-type approximation give an adequate description?

This issue has been investigated by 
Tamura et al. (1996; 2000) and Tamura (1998),
through detailed X-ray studies of Abell~1060.
Being a prototypical weak-CCC object 
with a nearly isothermal ICM as shown in subsection 2.3,
this cluster is also known to have little CEE (Jones, Forman 1984).
However, the ROSAT PSPC radial brightness profile exhibits a weak 
yet significant excess above a $\beta$-model (Tamura 1998; Tamura et al. 2000),
and hence cannot be explained by emission from 
an isothermal ICM confined in a King-type potential. 

In order to reconcile the good isothermality indicated by the ASCA data
and the weak CEE found in the ROSAT PSPC data,
Tamura (1998) and Tamura et al. (2000) have resorted to employ, 
instead of a King-type model, a total mass density distribution of the form
\begin{equation} %eq.8
\rho(R) = \rho_0 \left( \frac{R}{R_{\rm s}} \right)^{-\zeta}
                \left( 1+ \frac{R}{R_{\rm s}} \right)^{\zeta-3}~~,
\label{eq:NFW}
\end{equation}
where $\rho_0$, $R_{\rm s}$, and $\zeta$ are positive parameters.
This distribution, found by $N$-body simulations
(Navarro et al. 1996; Fukushige, Makino 1997; Moore et al. 1998; Ghigna et al. 2000),
has a singular cusp at the center,
but the implied gravitational potential remains finite there
as long as $\zeta < 2$.
For $\zeta=1$, this formula reduces to the ``universal halo'' profile 
of Navarro et al. (1996).
Then, the overall ASCA and ROSAT data of Abell~1060 have been reproduced successfully 
in terms of emission from a nearly isothermal ICM confined in the potential corresponding 
to equation (\ref{eq:NFW}) with $\zeta = 1.5 \pm 0.1$,
while the King-model profile has been ruled out (for detail, see Tamura et al. 2000).
Thus, the X-ray data of this prototypical non-cD cluster suggest
the gravitational potential with a central cusp,
%%%%%% added
and the derived value of $\zeta$ is in a very good agreement with 
that predicted by the $N$-body simulation (Ghigna et al. 2000).
%%%%%%%%%%%%%%%%%%%%% 

Similarly, Markevitch et al. (1999) reported
that the mass profiles of Abell~2199 and Abell~496 can be expressed
well by equation (\ref{eq:NFW}).
In addition, by analyzing a large number of objects observed with {\it ASCA},
Sato et al. (2000) concluded that equation (\ref{eq:NFW}) with $\zeta=1$
can successfully reproduce their dark halo shapes:
however, details are not given in their paper.

%----------------------- 3.5 ----------------------------
\subsection{The two types of the gravitational potential shape}
%----------------------- 3.5 ----------------------------

How does the cuspy potential profile, 
indicated by $N$-body simulations and observed from Abell~1060 (subsection 3.4),
relate to the hierarchically nested halo-in-halo type 
potential profile (subsection 3.3) found among cD clusters?
Observationally, the two phenomena are qualitatively similar, 
because they both imply an excess gravitating mass 
at the cluster center as noted in subsection 3.2.
Actually, the hard X-ray CEE of the Centaurus cluster can be 
explained by either of the two potential models (Ikebe et al. 1999),
because the strong CCC makes the potential shape ambiguous.
The case of Abell~2199 may be similar, 
as its potential shape can be expressed by
either the double-$\beta$ model (Xu 1998) 
or the cuspy model (Markevitch et al. 1999).
Conversely, the weak CEE of Abell~1060 may be described 
alternatively by a double-$\beta$ model.

%%%% added %%%%%%
From physics viewpoint, 
comparison of the central excess mass found in subsections 3.2 and 3.3 
against the cD's stellar mass yields a baryon faction of $0.2-0.3$,
at least in Fornax (Ikebe et al. 1996), 
Centaurus (Ikebe et al. 1999), and NGC~4636 (Matsushita et al. 1998).
Therefore, the central excess mass is still dark,
implying that the dark matter is clustered on the two distinct spatial scales.
This concept is essentially the same as the physics 
behind the formation of the cuspy potential,
that smaller-scale dark matter halos formed in earlier epochs
survive the subsequent hierarchical merging,
and remain as subhalos near the cluster center
(Navarro et al. 1996; Dubinsky 1998; Ghigna et al. 2000).
%%%%%%%%%%%%%%%%%%%%%%%%%%%%%

\begin{figure}[htb]
\centerline{\psfig{file=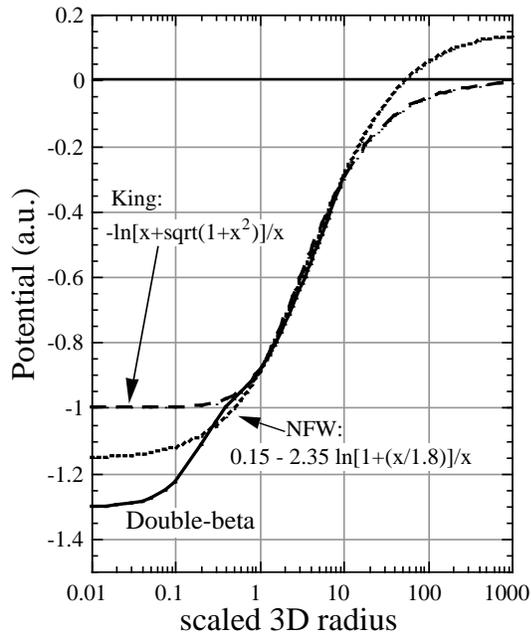,width=7cm}}
\caption{Schematic illustration of 
three typical gravitational potential models for clusters,
shown against the rescaled three-dimensional radius $x$.
The King-type model has the core radius corresponding to $x=1$.
The hierarchical double-$\beta$ potential assumes
that the ICM emissivity is given by a sum of 
two $\beta$-model components as expressed by equation (15),
with the two core radii corresponding to $x=1.0$ and $x=0.12$;
the two components are assumed to have the same temperature and the same $\beta=2/3$,
and the narrower emissivity component has a normalization
40 times as high as that of the wider emissivity component.
This simulates the Fornax cluster.
The Navarro-Frenk-White potential refers to equation (16), 
with $\zeta=1$ and $R_{\rm s}$ corresponding to $x=1.8$;
its normalization has been rescaled and offset
to match the other two profiles over a range of $x=1-5$.
The models are no longer accurate for $x>5$.
}
\label{fig:potential}
\end{figure}

%%%% modified %%%%%%%%%%%%%%%%%%%%%%%%%%%%%%%%%%%
In contrast to the above arguments for the close resemblance
between the cuspy and hierarchical potential models,
some observational facts argue against their identification.
First of all, the ASCA data of the Fornax cluster 
reveals the two distinct spatial scales without ambiguity.
Similarly, the hierarchical structure seen in NGC~4636 can be derived from the 
surface brightness profile in an non-parametric manner (Matsushita et al. 1998),
using Monte-Carlo simulations instead of the double-$\beta$ modeling.
Furthermore, the mass curve of Abell~1795 is better interpreted 
as exhibiting a hierarchical structure than a cuspy potential.
According to Williams et al. (1999),
the cluster core mass estimated through gravitational lensing
tends to exceed those predicted by the $N$-body simulations,
presumably due to the excess mass associated with the cD galaxy.
We therefore regard the two potential structures as distinct. 
In fact, as illustrated in figure~\ref{fig:potential}, 
the hierarchical potential can produce a much stronger CEE 
than the cuspy potential can (Makino et al. 1998; Tamura et al. 2000).
These implications are independent of our 2T formalism.
The difference between the two potential shapes may be consistent 
with the statistical result by Fujita, Takahara (1999),
that clusters form a two-parameter family,
with the two parameters being the system mass
and the degree of central mass concentration.
%%%%%%%%%%%%%%%%%%%%%%%%%%%%%%%%%%%

%%%% modified %%%%%%%%%%%%%%%%%%%%%%%%%%%%%%%%%%%
One likely scenario is that the central cusp is 
formed essentially in all clusters by the dark matter distribution, 
whereas the cusp develops into a much deeper central dimple 
as the cD galaxy builds up through, presumably, 
under a considerable baryonic energy dissipation,
as evidenced by the factor 2--3 higher baryon fraction
in the central dimple region than in the entire cluster.
To assess such a possibility,
in figure \ref{fig:M(r)}b we draw a straight line
which indicates the scaling relation found by Navarro et al. (1996) 
between $R_{\rm max}$ and the total integrated mass contained within it,
where $R_{\rm max}$ is the radius 
at which the circular velocity of the halo becomes maximum. 
Although the observed data points lie by a factor 2--3 above the prediction,
they will lie closer to the line,
if we subtract the mass contribution from the underlying cluster component.
Therefore, the central excess mass itself may be 
regarded as a self-gravitating dark halo,
and a cD cluster can be regarded as a halo-in-halo system.
We expect that future numerical simulations,
fully taking into gas dynamics,
will be able to reproduce the hierarchical potential profile
found in cD clusters.
%%%%%%%%%%%%%%%%%%%%%%%%%%%%%%%%%%%%%%%%%%%%%%%%

%%%% added %%%%%%%%%%%%%%%%%%%%%%%%%%%%%%%%%%%%%%%
In any case, our measurements of the cluster potential profile
imply the presence of a particular structure in the 
dark matter distribution underlying each cD galaxy.
This disagrees with the scenario (e.g., Fabian 1994)
which describes cD galaxies as a result of gas condensation in CFs.
%%%%%%%%%%%%%%%%%%%%%%%%%%%%%%%%%%%%%%%%%%%%%%%%%%%%%%%%%%%%%

%==================== 4 ===============================
\section{ICM METALLICITY IN THE CLUSTER CENTER REGIONS}
%==================== 4 ===============================
Production, confinement, and transport of heavy elements are 
a third important aspect of the X-ray study of clusters of galaxies.
ASCA observations have for the first time enabled
systematic studies of spatial distributions of heavy elements in the ICM,
particularly iron and silicon.
We again find characteristic phenomena around cD galaxies.

%--------------- 4.1 ----------------
\subsection{Cluster-wide properties}
%--------------- 4.1 ----------------

Excluding the CCC regions,
the average iron abundance of ICM is 0.2--0.3 solar,
without significantly depending on the cluster richness (Fukazawa et al. 1998).
As a result, the total iron mass in the ICM of a cluster amounts 
to $(1-3) \times 10^{-3}$  of the total stellar mass therein 
(Tsuru 1992; Arnaud et al. 1992).
Equivalently, the iron mass to light ratio 
(IMLR; Ciotti et al. 1991; Renzini et al. 1993; Renzini 1997),
i.e. the iron mass in the ICM normalized to the stellar light,
becomes $ (0.5-3) \times 10^{-2}$ in the solar unit.
This amount of iron is indeed comparable to 
the total iron locked in the stellar interior (Renzini et al. 1993). 

Such a large amount of iron (and other heavy elements) in the ICM
must have been produced in the stellar interior of member galaxies, 
particularly ellipticals (Arimoto, Yoshii 1987),
and subsequently ejected into the intra-cluster space.
However, observations with Ginga (Awaki et al. 1991), 
ROSAT (Forman et al. 1993), BBXRT (Serlemitsos et al. 1993), 
and ASCA (Awaki et al. 1994; Loewenstein et al. 1994; Matsumoto et al. 1997) 
have failed to detect the expected trace of such metal ejection processes
around individual galaxies:
the ISM metallicity of X-ray luminous ellipticals is at most $\sim 1$ solar,
and those of X-ray fainter ones can be even lower (Matsushita et al. 1997, 2000).
While these values are consistent with those expected from stellar mass-loss,
there is not much room left for the metal enrichment of the ICM by supernova (SN) products.
This clearly indicates that the SN products have been removed quickly 
from individual galaxies, and transported into the intra-cluster space, 
via, e.g., ram-pressure stripping and energetic outflow.
Indeed, the iron mass now contained in the ISM of all member galaxies of a cluster
would sum up to make only a few percent of that contained in the ICM.

\begin{figure}[htb]
\centerline{\psfig{file=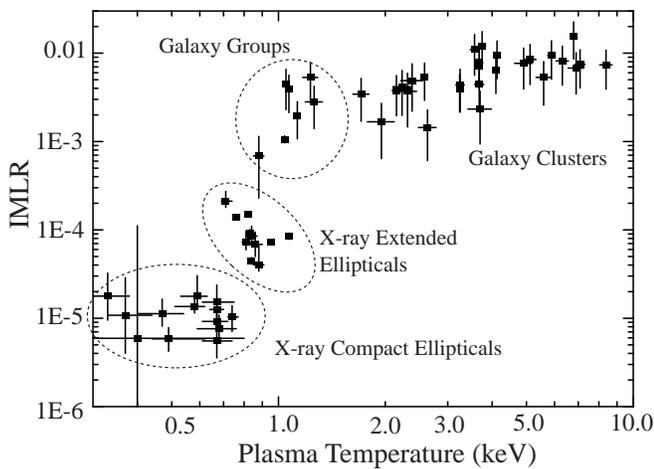,width=9cm}}
\caption{The IMLR as a function of the system mass.
Data for clusters and galaxy groups are taken from Fukazawa (1997),
where the IMLR is calculated within a radius 
where the ICM density falls below $3 \times 10^{-4}$ cm$^{-1}$.
Those for elliptical galaxies refer to Matsushita (1997),
where the IMLR is calculated with 4 times the optical effective radius.
}
\label{fig:IMLR}
\end{figure}

In order to reinforce this view,
in figure \ref{fig:IMLR} we summarize the IMLR of various objects,
as a function of their plasma temperature serving as a measure of the system richness.
The figure, originally devised by Fukazawa et al. (1996) and Ishimaru (1996, 1998),
includes clusters with different richness, galaxy groups, and elliptical galaxies.
Thus, the IMLR clearly decreases as the system gets poorer.
Evidently, poorer systems have lost most of the heavy elements produced in them,
presumably because they have too low an efficiency of gravitational confinement of the 
metal-enriched SN products (Fukazawa 1997; Matsushita 1997; Fukazawa et al. 2000).
The heavy elements are thus inferred to be escaping 
extensively from objects of lower hierarchy,
and the escaped materials enrich the systems of higher hierarchy.

In figure \ref{fig:IMLR}, elliptical galaxies form two distinct subgroups,
as discovered by Matsushita (1997) and Matsushita et al. (2000).
One class (``X-ray extended'' ones) comprises X-ray luminous objects such as NGC~4636, 
which also have higher ISM abundances.
They exhibit extended X-ray morphology 
indicative of larger-scale outer potential envelopes,
for which they act as mini-cD galaxies (subsection 3.3).
The other class  (``X-ray compact'' ones)  consists of  
ellipticals with low X-ray luminosities ($< 10^{40}$ ergs s$^{-1}$),
which lack outer potential envelopes.
The difference in IMLR between the two types can also be understood 
in the context of metal confinement and escape, 
because the outer potential envelopes associated with objects of the former class  
are though to improve the confinement efficiency of the energetic SN materials (Matsushita 1997).

The ASCA observations have for the first time yielded 
systematic measurements of the silicon-to-iron abundance ratio in the ICM,
which is a key to determining the relative importance of type Ia and type II supernovae.
In the outer regions of medium-richness clusters,
silicon has been found to be systematically over-abundant relative to iron 
by a factor of 1.5--2 in solar units
(e.g., Mushotzky et al. 1996; Tamura et al. 1996; Fukazawa et al. 1998).
This reveals the dominant role of type II SNe in the process of metal enrichment of ICM.
However, Fukazawa (1997) and Fukazawa et al. (1998) have discovered 
that the silicon over-abundance gradually disappears towards poorer clusters, 
and the poorest ones exhibit solar-like Si/Fe ratios
implying a significant contribution from type Ia SNe as well.
In other words, ``silicon mass to light ratio'' of the ICM decreases considerably,
e.g. by a factor of 2, from rich clusters to the poorest ones,
whereas the IMLR is approximately constant among clusters (figure \ref{fig:IMLR}).
As pointed out by Fukazawa et al. (1998),
one possible explanation of this effect is a selective escape of silicon;
presumably, silicon has been supplied early in the cluster evolution 
mainly in the form of galactic winds created by type II SNe,
which were energetic enough to escape from poorer systems.
In contrast, iron is likely to have been supplied by more prolonged activity 
of type Ia SNe without forming such energetic outflows. 

%----------new------ 4.2 -----------------------
\subsection{Large-scale radial behavior of IMLR}
%------------------- 4.2 -----------------------
Radial changes in the IMLR are expected to provide further clues
as to the metal production, confinement, and escape. 
For this purpose, we extend the concept of IMLR into 
``radial IMLR profile'', defined as 
\begin{equation}
\Phi(R) \equiv M_{\rm Fe}(R) / L_*(R)~~,
\end{equation}
where $M_{\rm Fe}(R)$ and $L_*(R)$ denote the radial profile of the 
iron mass in the ICM and that of the stellar light, respectively, 
both integrated within $R$ and expressed in solar units.
In outer regions of the cluster, $L_*(R)$ generally increases more slowly with $R$
than the integrated ICM mass profile, $M_{\rm ICM}(R)$.
Taking the AWM7 cluster for example, the differential forms of 
its $M_{\rm ICM}(R)$ and $L_*(R)$ can be approximated by 
$\beta$ models of $\beta = 0.58^{+0.03}_{-0.02}$ and $\beta \sim 0.8$, 
respectively (Ezawa et al. 1997).
Therefore, at large radii,
the ratio $M_{\rm ICM}(R)/L_*(R)$ increases as $\propto R^{0.66}$,  
where $0.66 = 3 \times (0.8-0.58)$ is the logarithmic slope difference
between $M_{\rm ICM}(R)$ and $L_*(R)$ in outer regions.
We then expect $\Phi(R) \propto R^{0.66} Z(R)$, 
where $Z(R)$ is the radial profile of the ICM iron abundance.

In many clusters, $Z(R)$ has so far been considered approximately constant 
as a function of $R$, except in the central regions.
However, Ezawa et al. (1997) have discovered 
that $Z(R)$ of AWM7 in fact decreases
over a cluster-wide spatial scale as $Z(R) \propto R^{-0.7 \pm 0.2}$,
outside $R \sim 4' = 120~h_{50}^{-1}$ kpc;
$M_{\rm Fe}$ is less extended than $M_{\rm ICM}$.
This effect just cancels out the outward increase in $M_{\rm ICM}(R)/L_*(R)$,
and makes $\Phi (R)$ almost radially constant at $ 2.8 \times 10^{-2}$.
Namely, $M_{\rm Fe}(R)$ of AWM7 behaves approximately proportional 
to $L_* (R)$ over spatial scales from $\sim 100$ kpc to $\sim 1$ Mpc,
implying that the iron density in the ICM faithfully traces the stellar light distribution. 

Similar results have been obtained with ASCA 
from the Perseus cluster (Ezawa 1998),
Abell~4059 (Kikuchi e al. 1999),
and Abell~2029 (Molendi, De Grandi 1999) as well. 
We presume that these effects,
i.e., the large-scale outward decrease in $Z(R)$
and the associated constancy of $\Phi (R)$ outside $R \sim 100$ kpc, 
are generally present in many clusters.
The reason why the effects have so far been observed from the four particular 
clusters may be ascribed to their high X-ray surface brightness 
and large angular extent.
With the future instrumentation,
we hence expect this effect to be detected ubiquitously from many clusters.
%%%%%%%%% added %%%%%%%%%
Actually, the XMM-Newton observation of the southern cluster 
S\'ersic 159-03 clearly reveals a large-scale ICM metallicity gradient
over a radial scale of $\sim 170$ kpc (Kaastra et al. 2001).
%%%%%%%%%%%%%%%%%%%%%%%%%%%%%%%%%%%%%%%%%%%%%%%%%%%%%%%%%%%%%%%

These results provide one of the first direct confirmations of the 
general consensus that the heavy elements in the ICM were indeed ejected 
by the cluster member galaxies into the intra-cluster space.
Furthermore, the constant IMLR profiles suggest
that the metals do not travel over a large distance (Ezawa et al. 1997),
even though they must be removed quickly from their source galaxies (subsection 4.1),
via, e.g., ram pressure stripping and energetic outflow.
Presumably, the galaxies have been swimming in the ICM 
while continuously depositing metal-enriched gas
(e.g., Charlton, Salpeter 1989).
The metals then ``mixed'' into the ICM and soon became hydrostatic,
resulting in a radial distribution 
which is similar to the galaxy distribution in a statistical sense.

We must remark here that the observations do not necessarily exclude 
homologous expansion or contraction in $M_{\rm ICM}(R)$ or $L_*(R)$,
in such a way that their core radii change but their $\beta$'s remain unchanged.
In these cases, $\Phi (R)$ would vary over the core region,
but would still remain constant outside it.
This urges us to study the metallicity behavior 
in the cluster center ($< 100$ kpc), in the following two subsections.

%------------------ 4.3-----------------------
\subsection{ICM metallicity near the cluster center}
%------------------ 4.3 ----------------------

The ICM metallicity has been observed to increase significantly 
toward the center of many cD clusters, including
Virgo (Koyama et al. 1991; Matsumoto et al. 1996), 
Centaurus (Fukazawa et al. 1994; Ikebe 1995; Ikebe et al. 1999), 
AWM7 (Xu et al. 1997), Abell~496 (Hatsukade et al. 1998),
Abell~262 (David et al. 1996),
Abell~4059 (Kikuchi et al. 1999),
and possibly several more objects (Ohashi et al. 1994).
%%%%%%%%%% Added %%%%%%%%%%%%%%%%%%
We can also add recent results from Chandra on the Hydra-A cluster (David et al. 2001)
and XMM-Newton measurements of Abell~1795 (Tamura et al. 2001).
%%%%%%%%%%%%%%%%%%%%%%%%%%%
This effect is limited to the central $\sim 100$ kpc scale
where the CCC and CEE phenomena are also observed,
and much more localized to the center
than the outward decrease in $Z(R)$ described in subsection 4.2.
Although these results are derived assuming
the two ICM phases to have the same abundances,
the metallicity increase in the hot phase is a robust result,
as shown by the enhanced Fe-K lines
in the central regions of most of these objects.
Furthermore, the CCC of the Centaurus cluster is 
at least as metal enriched as the central hot phase (Ikebe et al. 1999).
Therefore, we presume that the central increase in $Z(R)$
in these objects occurs in both hot and cool phases.

In contrast to these cD clusters, Abell~1060, the typical non-cD cluster,
exhibits an essentially constant $Z(R)$
from the center up to $\sim 250~h_{50}^{-1}$ kpc (Tamura et al. 1996).
The statistical study of 40 clusters by Fukazawa (1997),
described in subsections 2.2 and 2.3, 
confirm that this can be regarded as a 
systematic difference between cD and non-cD clusters:
the central $\sim 100$ kpc region of cD clusters exhibit 
systematically higher abundances than their outer regions,
while non-cD clusters exhibit spatially uniform ICM abundances over the core region.
(In the peripheral regions, both types of clusters are expected
to exhibit the large-scale abundance decrease described in subsection 4.2.)
We therefore conclude that the ICM abundance generally increases
within $\sim 100$ kpc of cD galaxies.

\begin{figure}[htb]
\centerline{\psfig{file=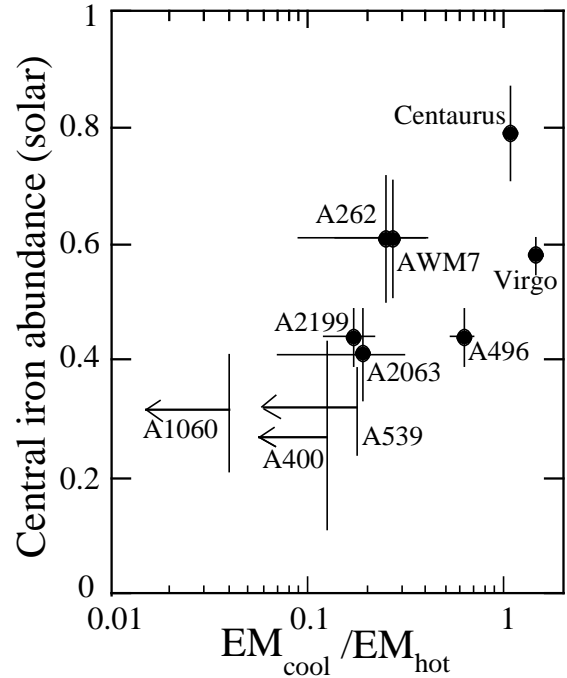,width=8cm}}
\caption{The ICM iron abundance in the cluster core region,
plotted as a function of the cool-to-hot emission integral ratio there (Table~1).
This updates Tamura et al. (1997).}
\label{fig:Z.vs.Qc/Qh}
\end{figure}

The above conclusion is reinforced by figure \ref{fig:Z.vs.Qc/Qh},
where we plot the central iron abundance against 
the $Q_{\rm c}/Q_{\rm h}$ ratio after Tamura et al. (1997).
This results is based on a subsample of table 1 with redshifts $<0.04$,
because the central metal-enriched regions of more distant objects
are difficult to resolve with ASCA.
Thus, the two quantities exhibit a tight positive correlation, 
with Centaurus and Virgo at one end of the distribution 
while the non-cD clusters (Abell~1060, Abell~400, and Abell~539) at the other end.
Since the CCC is a clear signature of cD galaxies (subsections 2.4 and 2.6),
the correlation indicates a close relation 
between the presence of a cD galaxy and the central metallicity enhancement.
A particular advantage of figure \ref{fig:Z.vs.Qc/Qh} is 
that its implication is independent of the definition of cD and non-cD clusters.

In a typical cD cluster, the excess iron mass contained 
in the central region (hot plus cool phases) amounts to  
$\sim 7 \times 10^9~M_\odot$ or less  (Fukazawa et al. 2000).
This can be supplied over the whole lifetime of a single giant
elliptical galaxy via type Ia SNe (Renzini et al. 1993).
Therefore, the excess metals at the cluster center can most 
naturally be regarded as a product of the cD galaxy,
as argued by several authors (e.g., Fukazawa et al. 1994, 2000; 
Ikebe et al. 1999; Kikuchi et al. 1999; David et al. 2001).
The ejected metals are thought to have mostly remained in the cluster core region,
because the cD galaxy is free from ram-pressure stripping effects
unlike the other galaxies that are moving through the ICM.
This viewpoint is reinforced by the systematic difference seen in the ICM 
chemical composition between central and outer regions of cD clusters 
(Fukazawa 1997; Fukazawa et al. 2000). 
Namely, in the metal-enriched cluster center regions
(again assuming the hot and cool phases to have the same abundances),
the Si/Fe ratio is approximately unity in solar units on average,
even though the silicon over-abundance is observed from outer regions
particularly when the cluster is relatively rich (subsection 4.1).
%%% Added %%%
The same conclusion has been derived by David et al. (2001) 
on the Hydra-A cluster using Chandra.
%%%%%%%%%%%%%%%%%%%%%
This solar-like Si/Fe ratio resembles those measured in the ISM of X-ray 
luminous elliptical galaxies (Matsushita 1997; Matsushita et al. 1998, 2000),
and suggests a significant contribution from type Ia SNe in the cD galaxy.
This particular result, considered to hold both for the hot and cool phases,
also gives a convincing support to our view presented in subsection 2.6
that the CCC is associated with the cD galaxy.

In contrast to cD clusters, 
the central regions of non-cD clusters are usually 
populated with several galaxies of comparable luminosities.
Since none of such galaxies is at rest in the gravitational potential,
they are all subject to the ram-pressure stripping,
and metals they produced must have been transported efficiently 
to the intra-cluster space just like those from the other member galaxies.
Furthermore, the relative motion of these galaxies
will efficiently mix the metal-rich ejecta into the ICM.
These ideas, already pointed out by Tamura et al. (1996),
give a natural account of the spatially uniform abundances of non-cD clusters.
This mechanism may apply even to some cD clusters.
For example, Tamura et al. (1996) argue 
that $Z(R)$ of the Fornax cluster stays rather constant at the center,
because its cD galaxy, NGC~1399, has a close companion NGC~1404
which must be in a strong interaction with NGC~1399.

Using ROSAT images and spatially averaged ASCA spectra,
Allen, Fabian (1998) showed 
that the emission-weighted ICM abundances of ``cooling-flow (CF)'' clusters 
are systematically higher than those of ``non cooling-flow (NCF)'' objects.
The sample they used (rather distant objects) is mostly disjoint from ours,
and their subsample classification is different from ours (cD vs. non-cD).
Nevertheless, among their 21 ``CF'' clusters,
five have Bautz-Morgan (B-M) type I,
one has type I-II, and four have type II; 
none is classified as B-M type II-III or III,
while the B-M type is unavailable for 10 objects.
On the contrary, of their nine ``NCF'' clusters,
one has B-M type II, three have type II-III, four have type III, and one unknown,
but none is type I or I-II.
Therefore, the Allen \& Fabian's classification is nearly identical to ours.
Their ``CF'' clusters in fact contain cD galaxies,
and the associated central metallicity enhancements, augmented by the CEE, 
presumably made the emission-weighted metallicity higher than those of ``NCF'' ones.
Thus, their results are consistent with ours,
but we consider the ICM metallicity to be more closely related to the 
presence/absence of a single dominant galaxy at the cluster center,
rather than to the CF strength.

In an attempt to reconcile the abundance increase 
of the Centaurus cluster with the CF hypothesis,
Reisenegger et al. (1996) argued that the metals ejected from the cD 
galaxy are swept back by CF to become compressed at the center.
They hence predict an anti-correlation between the central increment in $Z(R)$ and the CF strength, 
because the metals would drop out of the ICM as the cooling proceeds.
However, what we observe in figure \ref{fig:Z.vs.Qc/Qh} is quite opposite 
to the prediction, and rules out their interpretation.
Therefore, radiative metal dropouts are unlikely to be playing a major role.

%--------------- 4.4 --------------------------
\subsection{IMLR profiles at cluster centers}
%--------------- 4.4 ---------------------------

\begin{figure}[htb]
\centerline{\psfig{file=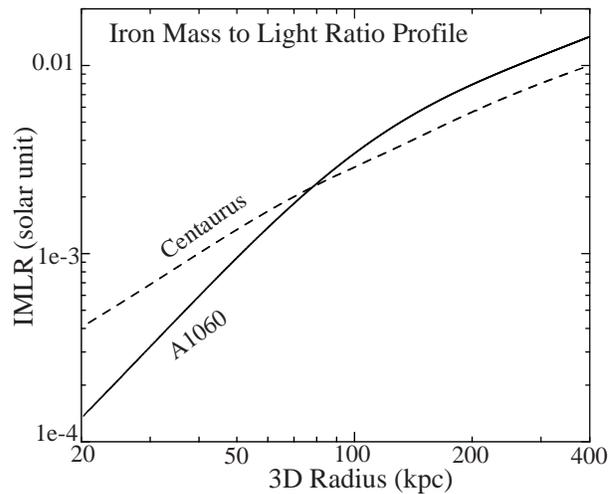,width=8cm}}
\caption{Radial profiles of the IMLR (iron mass to light ration) for 
the Centaurus cluster (dashed curve, from Ikebe et al. 1999)
and Abell~1060 (solid curve, from Tamura et al. 2000).
The calculation properly takes into account the angular response of ASCA.
}
\label{fig:IMLR(R)}
\end{figure}

How does $\Phi (R)$ behave near the center of clusters? 
For this purpose, in figure~\ref{fig:IMLR(R)}
we show $\Phi(R)$ of Abell~1060, calculated from Tamura et al. (2000).
Thus, $\Phi (R)$ clearly decreases to the center.
This is a trivial consequence of the two well established facts,
that the stellar light profile is more centrally peaked than the ICM mass profile,
and that the ICM metallicity of Abell~1060 is spatially uniform (subsection 4.3).

The central decrease in $\Phi (R)$ might be a result of cooling dropouts of the metals.
To examine this possibility, in figure \ref{fig:IMLR(R)} we have plotted 
$\Phi (R)$ of the Centaurus cluster as well, taken from Ikebe et al. (1999).
Although we again observe a central drop in the IMLR profile,
it is milder than that of Abell~1060,
because of the strong metallicity increase at the center.
Thus, the Centaurus cluster that has a much higher CF rate
exhibits a less marked decrease in $\Phi (R)$ than Abell~1060 that has little CF.
Therefore, the radiative cooling effects cannot be the main cause 
of the decrease in $\Phi (R)$,
in agreement with our inference made in subsection 4.3.
Incidentally, the curves of $\Phi (R)$ in figure~\ref{fig:IMLR(R)} are not affected 
significantly by the insufficient angular resolution,
because they have both been calculated by fully taking into account 
the instrumental point spread function.

The central decrease in IMLR is observed from other clusters as well.
For example, at a representative radius of $R \sim 100\ h_{50}^{-1}$ kpc, 
AWM7 and Abell~4059 exhibit IMLR of $3.2 \times 10^{-3}$ (converted from Xu et al. 1997)  
and $4 \times 10^{-3}$  (Kikuchi et al. 1999), respectively.
Both these values are very close to those of Abell~1060 and the Centaurus cluster.
From this result and those obtained in subsection 4.2, 
we can generally characterize $\Phi(R)$ by two features,
namely its constancy outside $100 \sim 200$ kpc, 
and its decrease toward the center.
We may restate the idea
that $L_*(R)$ and $M_{\rm Fe}(R)$ share nearly the same value of $\beta$
which in turn is larger (steeper) than that for $M_{\rm ICM}(R)$,
while $L_*(R)$ has a smaller core radius than $M_{\rm Fe}(R)$.

Because the ejected metals are likely to be conserved in the ICM,
the suggested difference in the core radius,
between $L_*(R)$ and $M_{\rm Fe}(R)$, 
is thought to reflect time-dependent changes 
in the spatial distributions of the stars and/or metals.
More specifically, the phenomenon can be explained as a result of 
either a radial contraction in the stellar distribution,
or a radial expansion in the metal distribution.
The latter mechanism must be operating to a certain extent,
because we have invoked repeatedly the metal escape effects.
However, the uniform ICM abundances at the center of non-cD clusters would require
that the suggested metal outflow is balanced in detail 
by a radial expansion of the primordial ICM.
Since such a fine tuning is generally unlikely,
we consider the former mechanism, i.e. the galaxy infall, 
to be more dominant. 
The galaxies, while continuously ejecting metals,
are therefore suggested to have gradually fallen to the cluster center,
or merged into bigger ones residing in central regions,
to achieve the centrally-peaked distribution of the stellar component.

The values of IMLR so far attributed to the cluster core regions, 
several times $10^{-3}$,
all refer to the iron contained in both hot and cool phases.
In contrast, if we consider the CCC only,
the IMLR further reduces to several times $10^{-4}$.
This is comparable to those of the ISM of X-ray luminous ellipticals.
This gives further support to our conclusion made in subsection 2.6 
that the CCC is the cD's ISM.
Presumably, a major fraction of the iron produced in the cD galaxy 
is contained in the hot phase in the form of a central metal excess,
which represents what would have normally been removed 
if the galaxy were moving through the ICM.

\vspace{-2mm}
%============  ========
\section{DISCUSSION}
%============5  ========

%------------------ 5.1 --------------
\subsection{Summary of new results}
%------------------ 5.1 --------------

So far, many cluster X-ray investigators purposely 
avoided analyzing the data from cluster central regions,
except when they try to develop the CF scenario.
We have challenged this issue through ASCA observations of nearby clusters, 
and arrived at a new picture 
of the gas and mass structure of the central regions of clusters.
Our results are summarized in the following three aspects.

In section 3, we have found 
that the gravitational potential exhibits a marked deepening at the cluster center,
as evidenced by the presence of CEE in hard X-rays.
More specifically, the gravitating mass distribution can be described by 
either a hierarchical model involving two characteristic scale lengths
(corresponding to the cluster and the cD galaxy),
or a scale-free distribution with a central cusp as predicted by numerical simulations.
We tentatively assign the former type of distribution to cD clusters,
and the latter to non-cD ones,
although the sample is still small.
%%%% added %%%%%
In particular, we emphasize the halo-in-halo structure formed around each cD galaxy.
%%%%%%%%%%%%%%%%%%%%%%%%%%%%

The study in section 2 employing the 2T formalism has lead us to propose
that the CCC, which is specific to cD clusters, 
%%%% changed %%%%%
has characteristics of the ISM (inter-stellar medium) of the cD galaxy filling the cD's own potential,
rather than of a cooling portion of the ICM.
%%%%%%%%%%%%%%%%%%%%%%%%%%%%
To the five supporting facts listed in subsection 2.6,
we can add the two metallicity arguments presented in subsection 4.3,
that the regions around cD galaxies exhibit metallicity enhancements,
and that the chemical compositions there
somewhat differ from those of the bulk ICM.
The effect of ICM cooling, if any, 
turns out to be much smaller than was thought previously,
suggesting that significant ICM heating mechanisms are operating.

Our study of the ICM metallicity in section 4 
incorporating the concept of the IMLR profile
have revealed two interesting inferences.
One is that the metals have been escaping extensively from a poor system,
a part of which can be trapped by the potential of a surrounding system of a higher hierarchy.
The other, though more speculative, is 
that the radial galaxy distribution has been 
gradually shrinking relative to the ICM distribution, 
and possibly relative to the dark-matter distribution as well.
The observed decrease in the IMLR profile toward the center 
may be explained by a combination of these two effects.

As a straightforward application of our results,
we can for the first time solve the long-lasting confusion 
as to the nature of the bright X-ray emission associated with M87, 
the cD galaxy of the Virgo cluster.
So far, some authors (e.g., Takano et al. 1989; B\"ohringer et al. 1994) 
interpreted it as a part of the Virgo ICM emission,
while others (e.g., Fabricant, Gorenstein 1983; Beuing et al. 1999)
discussed it in terms of the ISM emission from M87 as an elliptical galaxy.
%%%% changed
According to our picture,
%%%%%%%%%%%%%%%%%%%%%%%%%%%%%%%%%%
the central cool phase corresponds to the ISM associated with M87,
whereas the hot phase around M87 represents the Virgo ICM 
of which the density is enhanced by the potential drop associated with M87.
This statement, first suggested by Matsumoto et al. (1996),
concisely summarizes the present work.

%%%% Added 
A large amount of new data are accumulating from Chandra and XMM-Newton observations.
Although detailed comparison of these results with ours is yet to be carried out,
the main points we have emphasized in this paper are
mostly being reconfirmed and reinforced by these two powerful missions.
%%%%%%%%%%%%%%%%%%%%%%%%%%%%%%%%%%

%----------------------- 5.2 -----------------------------------
\subsection{Heating and thermal stability of the cool component}
%----------------------- 5.2 -----------------------------------
As mentioned in subsection 2.6,
our scenario is subject to several theoretical issues yet to be solved;
(1) how the cool ISM phase is separated and thermally insulated from the hot ICM phase, 
against rapid heat conduction;
(2) what supplies the cool phase with the large amount of energy 
(up to $\sim 10^{44}$ ergs s$^{-1}$) necessary to sustain 
the X-ray radiation and prevent it from the thermal collapse;
and
(3) how the heating balances the cooling in a stable manner.
Referring to Makishima (1997ab, 1999ab) and Ikebe et al. (1999),
we below present some speculative ideas that might answer these questions.

The ICM is generally magnetized 
up to a few micro Gauss level (e.g., Kronberg 1994; Eilek 1999).
The magnetic field is known to be particularly strong
near cD galaxies, reaching $10-100~\mu$G (Taylor et al. 1999).
As a natural consequence, we may invoke magnetic fields
as the required thermal insulator between the two phases
(Makishima 1994b, 1997b, 1999a).
Even though the actual field strengths may fall
below the equipartition value ($\sim 30~\mu$ G in typical cases),
the heat conduction can be suppressed by many orders of magnitude.
This could solve the first question.

As to the second question, various heating source candidates have long been considered,
including supernovae, active galactic nuclei, drag due to the galaxy motion, and so on.
A general consensus is
that none of these candidates provide heating luminosity high enough to sustain 
the radiative energy loss of the cool phase (e.g., Bregman, David 1989; Fabian 1994).
However, one important possibility has not been fully considered;
i.e., magnetohydrodynamic (MHD) effects 
as the member galaxies move through the ICM (Makishima 1997ab,1999b).
Because the ICM is such an ideal classical plasma 
with extremely high magnetic Reynolds numbers,
we expect the galaxy motion to cause significant MHD turbulence 
and frequent magnetic reconnection (Norman, Meiksin 1996).
Then, the kinetic energy of galaxies may be dissipated
with a much higher efficiency 
than would be expected when the gas is neutral.
The recently reported X-ray ``wakes'' (Drake et al. 2000), 
based on the ROSAT HRI image of Abell 160, may provide evidence for such effects.
Although calculation of the heating luminosity of the 
proposed mechanism is subject to large uncertainties, 
a crude estimate suggests that it might work (Makishima 1999b).
%Then, is the heating efficient enough?
%Supposing that a typical cluster contains 100 galaxies 
%with a velocity dispersion of $v=700$ km s$^{-1}$,
%the heating luminosity due to the galaxy drag becomes
%$L_{\rm drag} \sim 100\ \xi n m_{\rm p} v^3 S
%\sim 2 \times 10^{45} \xi$ ergs s$^{-1}$ (Makishima 1999b).
%Here $\xi$ is the efficiency of thermalization of the transferred kinetic energy,
%$n \sim 3 \times 10^{-3}$ cm$^{-3}$ is the ICM density,
%and $S=\pi r^2$ with $r \sim 20$ kpc is the MHD cross section of each galaxy.
%Although $\xi$ is difficult to estimate, a moderately low value, e.g. $\sim 1$\%,  
%would yield $L_{\rm drag} \sim 2 \times 10^{43}$ ergs s$^{-1}$.
%The available energy is sufficient; 
%the typical total stellar mass of $1 \times 10^{13}~ M_\odot$ in a cluster
%implies a total kinetic energy of $5 \times 10^{61}$ ergs,
%and the above-estimated $L_{\rm drag}$ can be sustained
%if $\sim 30$\% of this energy is released over the Hubble time.

If the proposed mechanism invoking the MHD effects is actually operating,
we should expect reactions from the ICM to the member galaxies.
There are in fact several hints of such effects in the existing data.
One is the suggestion made in subsection 4.4,
that the radial galaxy distribution may have been shrinking relative to that of the ICM.
It might be that the galaxies have lost some portion of 
their kinetic energies through interaction with the ICM,
and have gradually fallen inwards, 
while the ICM expanded by receiving the released energy.
The observed increase in the fraction of elliptical galaxies toward the 
cluster center (e.g., Whitmore, Gilmore 1993) can be another piece of evidence,
because the drag force will accelerate mergers
of smaller spirals into larger ellipticals.
Similarly, the observed galaxy ``metamorphosis'' 
from distant to nearby clusters (Dressler et al. 1994) could be a result 
of enhanced galaxy interactions and mergers through the ICM drag (Makishima 1999a). 
 
In addition to the proposed galaxy-ICM interaction,
we can speculate on another ICM heating mechanism localized around each cD galaxy,
where the heating energy is most vitally needed.
As described in section 3 and figure \ref{fig:M(r)}b,
a cD galaxy is likely to form a self-gravitating core,
immersed in the larger cluster halo.
The self-gravitating energy reaches
$\sim G (\delta M)^2/R_{\rm IF} \sim 1 \times 10^{62}$ ergs,
where $\delta M$ refers to equation~(\ref{eq:dMnum}) 
and $R_{\rm IF} \sim 100$ kpc is the typical interface radius (subsection 3.3).
As the self-gravitating core gradually shrinks, 
possibly under a significant baryonic influence,
half the released gravitational energy would be radiated,
while the remaining half would be spent in the ICM heating.
If, e.g.,  $\sim 10\%$ of the self-gravitating energy 
is thus released over the Hubble time,
the available heating luminosity, $\sim 1 \times 10^{43}$ ergs s$^{-1}$,
may be sufficient to prevent the ICM from radiative collapse (Makishima 1999b).

Even though the heating energy may be available,
the volume cooling rate of the ICM due to radiation is 
proportional to $n^2$ as given by equation~(\ref{eq:emissivity}),
while the volume heating rate is normally proportional to $n$.
Therefore, it is usually difficult for the heating mechanisms to stably balance the cooling.
However, the solar corona, confined within many magnetic loops,
is thermally stabilized by the self-regulating 
Rosner-Tucker-Vaiana (1978) mechanism (Kano, Tsuneta 1995).
As an analogy, we speculate that the magnetic field in the cluster core region
takes a form of numerous loops anchored to the cD galaxy;
the loops are surrounded by the hot phase,
and their interior is occupied by the cool phase.
Makishima (1997b) termed this concept ``cD corona'':
such structured fields are suggested in the literature,
(Ge, Owen 1994; Taylor et al. 1994; Owen et al. 1999; B\"ohringer 1999).
If so, the heating and cooling could be stably balanced
by the Rosner-Tucker-Vaiana mechanism.
This idea might give an answer to the third question.

\vspace{-2mm}

%========== 6 =================
\section{Conclusion}
%========== 6 =================
By combining several observational works on 
central regions of galaxy clusters achieved with ASCA,
we have arrived at a novel view therein.
This view describes the region around a cD galaxy 
as a site of significant and active evolution,
where a large amount of heavy elements are produced,
a self-gravitating core develops,
and presumably certain ICM heating processes are operating.
The scenario makes a contrast to the previous view
which emphasized the role of radiative plasma cooling.
The recent Chandra and XMM-Newton results clearly favor our scenario.

\vspace{4mm}
%{Acknowledgements}
The authors thank all members of the ASCA team.
They are also grateful to Drs. S. Tsuneta, K. Shibata, and T. Terasawa
for helpful discussions.
We also thank the anonymous referee for careful reading of the manuscript
and helpful comments.
The present work is supported in part 
by the Grant-in-Aid for Center-of-Excellence, No. 07CE2002, 
from Ministry of Education, Science, Sports and Culture of Japan. 

%%%%%%%%%%%%%%%%%%%%%%%%%%%%%%%%%%%%%%%%
%\clearpage
\section*{References}
%%%%%%%%%%%%%%%%%%%%%%%%%%%%%%%%%%%%%%%%
\small
\re Abell G.O., Corwin H., Olowing R. 1989, ApJS 70, 1
\re Allen W., Fabian A. C. 1994, MNRAS 269, 409
\re Allen W., Fabian A. C. 1997, MNRAS 286, 583
\re Allen W., Fabian A. C. 1998, MNRAS 297L, 63
\re Anders E., Grevesse N. 1989, Geochim. Cosmochim. Acta 53, 197
\re Arimoto N., Yoshii Y. 1987, A\&A 173, 23
\re Arnaud M., Rothenflug, R., Boulade O., Vigroux L., Vangioni-Flam E.
    1992, A\&A 254, 49
\re Awaki H., Koyama K., Kunieda H., Takano S., Tawara Y., Ohashi T. 
    1991, ApJ 366, 88
\re Awaki H., Mushotzky R. F., Tsuru T., Fabian A. C., Fukazawa Y., Loewenstein M.,
    Makishima K., Matsumoto H. et al. 1994, PASJ 46, L65
\re Bautz L. P., Morgan W. W. 1970, ApJL 162, L149
\re Beuing  J., D\"obereiner  S., B\"ohringer  H., Bender  R. 
      1999, MNRAS, 302, 209
\re B\"ohringer H. 1999, in Diffuse Thermal and Relativistic Plasma in Galaxy Clusters,
    ed. H. B\"ohringer, L. Feretti, P. Sch\"ucker (Max-Planck-Institute f\"ur
    Extraterrestrische Physik) p115
\re B\"ohringer H., Briel U. G., Schwartz R. A., Voges W., Hartner  G., 
      Tr\"umper, J. 1994, Nature 368, 828
\re Bregman J. N., David L. P. 1989, ApJ 341, 49
\re Briel, U., Henry, J.P. 1996, ApJ 472, 131
\re Burke B. E., Mountain R. W., Daniels P. J., Dolat V. S. 
      1994, IEEE Trans. Nuc. Sci. 41, 375
\re Canizares C. R., Clark G. W., Jernigan J. G., Markert T. H. 1982, ApJ 262, 33
\re Canizares C. R., Fabbiano G., Trinchieri G. 1987, ApJ 312, 503
\re Cen R., Ostriker J. 1999, ApJ 514, 1    %Warm gas
\re Charlton J. C., Salpeter E. E. 1989, ApJ 346, 101
\re Ciotti L., D'Ercole A., Pellegrini S., Renzini A. 1991, ApJ 376, 380
\re David L. P., Arnaud K. A.,  Forman W., Jones C. 1990, ApJ 356, 32
\re David L. P., Jones C., Forman W. 1996, ApJ 473, 692
\re David L. P., Nulsen P. E. J., McNamara B. R., Forman W., Jones C.,
    Ponman T., Robertson B., Wise M. 2001, ApJ in press (astro-ph/0010244)
\re Drake N., Merrifield M. R., Sakelliou I., Pinkney J. C. 2000, MNRAS 314, 768 
\re Dressler A. 1979, ApJ 231, 659
\re Dressler A., Oemler A. Jr., Sparks W. B., Lucas\ R. 1994, ApJL 435, L23
\re Dubinski, J. 1998, ApJ 502, 141
\re Dwarakanath K. S., van Gorkom J. H., Owen F. N. 1994, ApJ 432, 496
\re Edge A. C., Stewart G. C., Fabian, A. C. 1992, MNRAS 258, 177 
\re Eilek J. 1999, in Diffuse Thermal and Relativistic Plasma in Galaxy Clusters,
    ed. H. B\"ohringer, L. Feretti, P. Schuecker (Max-Planck-Institute f\"ur
    Extraterrestrische Physik) p71
\re Ezawa H. 1998, PhD Thesis, University of Tokyo
\re Ezawa H., Fukazawa Y., Makishima K., Ohashi T., Takahara F., Xu H., Yamasaki N. Y. 
    1997, ApJL 490, L33
\re Fabian A. C. 1994, ARA\&A 32, 277
\re Fabian A. C., Arnaud K. A., Bautz M. W., Tawara Y. 1994, ApJL 436, L63
\re Fabricant D., Gorenstein P. 1983, ApJ 267, 535
\re Forman W., Jones C., David L., Franx M., Makishima K., Ohashi T.
    1993, ApJL 418, L55 
\re Forman W., Jones C., Tucker W. 1985, ApJ 293, 102
\re Fujita Y., Takahara F. 1999, ApJL 519, L51
\re Fukazawa Y. 1997, PhD Thesis, University of Tokyo
\re Fukazawa Y., Makishima K., Matsushita K., Yamasaki N. Y., Ohashi T., 
    Mushotzky R. F., Sakima Y., Tawara Y., Yamashita K. 1996, PASJ 48, 395
\re Fukazawa Y., Makishima K., Tamura T., Ezawa H., Xu H., Ikebe Y., Kikuchi K., 
    Ohashi T. 1998, PASJ 50, 187
\re Fukazawa Y., Makishima K., Tamura T., Nakazawa K., Ezawa Y., Ikebe Y.,
    Kikuchi K., Ohashi T. 2000, MNRAS 313, 21
\re Fukazawa Y., Nakazawa K., Isobe N, Makishima K., Matsushita K., Ohashi T., Kamae T.
    2000, ApJL, submitted
\re Fukazawa Y., Ohashi T., Fabian A. C., Canizares C. R., Ikebe Y., Makishima K.,
    Mushotzky R. F., Yamashita K. 1994, PASJ 46, L55
\re Fukushige T., Makino J. 1997, ApJ 477, 9
\re Ge J., Owen F. N. 1994, AJ 108, 1523
\re Ghigna S., Moore B., Governato F., Lake G., Quinn T., Stadel, J. 2000,
    ApJ 544, 616
\re Hatsukade I., Ishizaka J., Yamauchi M., Takagishi K. 1998,
    in The Hot Universe (Proc. IAU Symposium 188), 
    ed K. Koyama, S. Kitamoto, M. Itoh (Kluwer Academic, Dordrecht) p134
\re Huang Z., Sarazin L. 1998, ApJ 496, 728
\re Ikebe Y. 1995, PhD thesis, University of Tokyo
\re Ikebe Y., Ezawa H., Fukazawa Y., Hirayama M., Ishisaki Y., Kikuchi K., Kubo H.,
    Makishima K. et al. 1996, Naure 379, 427
\re Ikebe Y., Fukazawa Y., Tamura T., Makishima K., Ohashi T. 1997a, 
      in X-ray Imaging and Spectroscopy of Cosmic Hot Plasmas,
      ed F. Makino, K. Mitsuda (University Academy Press, Tokyo) p57
\re Ikebe Y., Makishima K., Ezawa H., Fukazawa Y., Hirayama M., Honda H.,
    Ishisaki Y., Kikuch K.  et al. 1997b, ApJ 481, 660
\re Ikebe Y., Makishima K., Fukazawa Y., Tamura T., Xu H., Ohashi T., Matsushita K. 
    1999, ApJ 525, 58
\re Ishimaru Y. 1996, PhD Thesis, University of Tokyo
\re Ishimaru Y. 1998, in Origin of Matter and Evolution of Galaxies 97,
      ed S. Kubono, T. Kajino, K. I. Nomoto, I. Tanihata
      (World Scientific, Singapore) p124
\re Johnstone R. M., Naylor T., Fabian A. C. 1991, MNRAS 248, p18
\re Jones C., Forman, W. 1984, ApJ 276, 38
\re Kaasttra J. S., Ferrigno C., Tamura T., Paerels F. B. S., Peterson J. R., Mittaz P. D.
    2001, A\&A 365, L99
\re Kaneda H. 1997, PhD thesis, University of Tokyo
\re Kaneda H., Makishima K., Yamauchi S., Koyama K., Matsuzaki K., Yamasaki N. Y. 1997,
     ApJ 491, 638
\re Kano R., Tsuneta S. 1995, ApJ 454, 934
\re Kikuchi K., Furusho T., Ezawa H., Yamasaki N., Ohashi T., Fukazawa Y., Ikebe Y.
      1999, PASJ 51, 301
\re Koyama K., Makishima K., Tanaka Y., Tsunemi H. 1986, PASJ 38, 121
\re Koyama K., Takano S., Tawara Y. 1991, Nature 350, 135
\re Kronberg P. P. 1994, Rep. Prog. Phys. 325
\re Lazzati D., Chincarini G. 1998, A\&A 339, 52 
\re Loewenstein M., Mushotzky R. F., Tamura T., Ikebe Y., Makishima K.,
      Matsushita K., Awaki H., Serlemitsos P. J. 1994, ApJL 436, L75 
\re MacKenzie M, Schlegel E.M., Mushotzky R. 1996, ApJ 468, 86
\re Makino N., Sasaki S., Suto Y. 1998, ApJ 497, 555
\re Makishima K. 1994a, in New Horizon of X-Ray Astronomy,
      ed F. Makino, T. Ohashi (Universal Academy Press, Tokyo) p171
\re Makishima K. 1994b, in Elementary Processes in Dense Plasmas,
      ed. S. Ichimaru, S. Ogata (Addison-Wesley, New York) p47
\re Makishima K. 1995, in Dark Matter (AIP Proceedings No.336),
      ed S. S. Holt, C. Bennet (AIP, New York) p172
\re Makishima K. 1996, in UV and X-ray Spectroscopy of Astrophysical 
      and Laboratory Plasmas, ed K. Yamashita, T. Watanabe
      (Universal Academy Press, Tokyo) p167 
\re Makishima K. 1997a, Plasma Phys. Control. Fusion 39, A15
\re Makishima K. 1997b, in Imaging and Spectroscopy of Cosmic Hot Plasmas,
      ed. F. Makino, K. Mitsuda (Universal Academy Press, Tokyo) p137
\re Makishima K. 1998, in The Hot Universe (Proc. IAU Symposium No.188), 
      ed K. Koyama, S. Kitamoto, M. Itoh (Kluwer Academic, Dordrecht) p181
\re Makishima K. 1999a, in Observational Plasma Astrophysics: Five years of
    Yohkoh and beyond, ed T. Watanabe, T. Kosugi, A. Sterling 
    (Kluwer Academic, Dordrecht) p51 
\re Makishima K. 1999b, Astronomische Nachrichten 320, 161
\re Makishima K., Tashiro M., Ebisawa K., Ezawa H., Fukazawa Y., Gunji S.,
    Hirayama M., Idesawa E. et al. 1996, PASJ 48, 171
\re Markevitch M., Vikhlinin A., Forman W. R., Sarazin C. L. 1999, ApJ 527, 545
\re Matilsky T., Jones C., Forman W. 1985, ApJ 291, 621
\re Matsumoto H., Koyama K., Awaki H., Tomida H., Tsuru T., Mushotzky R., Hatsukade I. 
    1996, PASJ 48, 201
\re Matsumoto H., Koyama K., Awaki H., Tsuru T., Lowenstein M., Matsushita K.
    1997, ApJ 482, 133
\re Matsushita K. 1997, PhD Thesis, University of Tokyo
\re Matsushita K., Makishima K., Ikebe Y., Rokutanda E., Yamasaki N. Y., Ohashi, T. 
    1998, ApJL 499, L13
\re Matsushita K., Makishima K., Rokutanda E., Yamasaki N. Y., Ohashi, T.
    1997, ApJL 488, L125
\re Matsushita K., Ohashi T., Makishima K. 2000, PASJ 52, 685
\re Merritt D., ApJ 289, 18
\re Mohr J. J., Mathiesen B., Evrard A. E. 1999, ApJ 517, 627
\re McNamara B. R., Wise M., Nulsen P. E. J., David L. P., Sarazin C. L., 
    Bautz M., Markevitch M., Vikhlinin A. et al. 2000, ApJL 534, L135
\re Molendi S., De Grandi S. 1999, A\&A, 351, L41
\re Moore B., Ghigna S., Governato F., Lake G., Quinn T., Stadel J, Tozzi P. 
    1999, ApJ 524, L19
\re Moore B., Governato F., Quinn T., Stadel. J., Lake G. 1998, ApJ 499, L5
\re Mulchaey J. S., Zabludoff A. I. 1998, ApJ 496, 73
\re Mushotzky R., Loewenstein M., Arnaud K. A., Tamura T., Fukazawa Y.,
    Matsushita K., Kikuchi K., Hatsukade I. 1996, ApJ 466, 686
\re Mushotzky R. F., Szymkowiak A. E. 1988, in Cooling Flows in Clusters and
    Galaxies (Kluwer Academic, Dordrecht) p53
\re Navarro J. F, Frenk C. S, White S. D. M. 1996, ApJ 462, 563
\re Neumann D. M., Arnaud M. 1999, A\&A 348, 711
\re Nulsen P. E. J. 1986, MNRAS 221, 377
\re Nulsen P. E. J. 1998, MNRAS 297, 1109
\re Ohashi T., Ebisawa K., Fukazawa Y., Hiyoshi K, Horii M., Ikebe, Y., Ikeda, H.,
    Inoue, H. et al. 1996, PASJ 48, 157
\re Ohashi T., Fukazawa Y., Ikebe Y., Ezawa H., Tamura T., Makishima K.
    1994, in New Horizon of X-Ray Astronomy, ed F. Makino, T. Ohashi
    (University Academy Press, Tokyo) p273
\re Owen F., Eilek J., Kassim N. 1999,
    in Diffuse Thermal and Relativistic Plasma in Galaxy Clusters,
    ed. H. B\"ohringer, L. Feretti, P. Schuecker 
    (Max-Planck-Institute f\"ur Extraterrestrische Physik) p107
\re Peterson J. R., Paerels F. B. S., Kaastra J. S., Arnaud M., Reiprich T. H.,
    Fabian A. C., Mushotzky R. F. et al. 2001, A\&A 365, L104
\re Pownall H. R., Steart G. C. 1996, in R\"ontgenstrahlung from the Universe,
      ed H. U. Zimmermann, J. Tr\"umper, H. Yorke (MPE Report 263, Garching) p609
\re Reisenegger A., Miralda-Escude J., Waxman E. 1996, ApJ 457, L11
\re Renzini A. 1997, ApJ 488, 35
\re Renzini A., Ciotti L., D'Ercole A., Pellegrini S. 1993, ApJ 419, 52
\re Rood H., Sastry G. 1971, PASP 83, 313
\re Rosner R., Tucker W., Vaiana G. 1978, ApJ 220, 643
\re Rothenflug R., Vigroux L., Mushotzky R. F., Holt S. S. 1984, ApJ 279, 53
\re Sarazin C. L. 1988, X-Ray Emission from Clusters of galaxies 
      (Cambridge University Press)
\re Sarazin C. L. 1997, in Galactic and Cluster Cooling Flows, 
      ed N. Soker (Publ. Astr. Soc. Pacific), p172
\re Sato S., Akimoto F., Furuzawa A., Tawara Y., Watanabe M. 2000, ApJ 537, L73
\re Schindler S., Hattori M., Neumann D.M., B\"ohringer H. 
      1997, A\&A 317, 646
\re Schombert J. M. 1986, ApJS 60, 603
\re Serlemitsos P. J., Loewenstein M., Mushotzky R. F., Marshall, F. E., Petre R. 
      1993, ApJ 413, 518
\re Singh K. P., Westergaard N. J., Schnopper H.W. 1988, ApJ 330, 620
\re Stewart G. C., Fabian A. C., Jones C., Forman W. 1984, ApJ 285, 1
\re Struble M., Rood H. 1987, ApJS 93, 1035
\re Suginohara T., Ostriker J. P. 1998, ApJ 507, 16
\re Takahara M., Takahara F. 1979, Prog. Theor. Phys. 62, 125
\re Takano S., Awaki H., Koyama K., Kunieda H., Tawara Y., Yamauchi S., 
      Makishima K., Ohashi T. 1989, Nature 340, 289 
\re Tamura T. 1998, Ph.D.Thesis, University of Tokyo
\re Tamura T., Day C. S., Fukazawa Y, Hatsukade I., Ikebe Y.,
    Makishima K., Mushotzky R. F., Ohashi T. et al. 1996, PASJ 48, 671
\re Tamura T., Ikebe Y., Fukazawa Y., Makishima K., Ohashi T. 1997, 
    in X-ray Imaging and Spectroscopy of Cosmic Hot Plasmas,
    ed F. Makino, K. Mitsuda (University Academy Press, Tokyo) p127
\re Tamura T., Kaastra J., Peterson J. R., Paerels F. B. S., Mittaz J. P. D.,
    Trudolyubov S. P., Stewart G. et al. 2001, A\&A 365, L87
\re Tamura T., Makishima K., Fukazawa Y., Ikebe Y., Xu H. 2000, ApJ 535, 602
\re Tanaka Y., Inoue H., Holt S. S. 1994, PASJ 46, L37
\re Taylor G. B., Allen S. W., Fabian A. C. 1999, 
    in Diffuse Thermal and Relativistic Plasma in Galaxy Clusters,
    ed. H. B\"ohringer, L. Feretti, P. Schuecker 
   (Max-Planck-Institute f\"ur Extraterrestrische Physik) p77
\re Taylor G., Barton E., Ge J. 1994, AJ 107, 1942
\re Thomas P. A., Fabian A. C., Nulsen P. E. J. 1987, MNRAS 228, 973
\re Trinchieri G., Fabbiano G. 1985, ApJ 296, 447
\re Trinchieri G., Kim D.-W., Fabbiano G., Canizares C. R. C. 1994, ApJ 428, 555
\re Tsuru T. 1992, PhD Thesis, University of Tokyo
\re Vikhlinin A., McNamara B.R., Hornstrup A., Quintana H., 
      Forman W., Jones C., Way M. 1999, ApJ 520, L1
\re White D. A., Fabian A. C., Johnstone R. M., Mushotzky R. F., Arnaud K. A. 
      1991, MNRAS 252, 72
\re Whitmore B., Gilmore D.M. 1993, ApJ 407, 489
\re Williams L. L. R., Navarro J.F., Bartelmann M. 1999, ApJ 527, 535
\re Wu X.-P., Chiueh T., Fan L.-Z., Xue Y.-I. 1998, MNRAS 301, 861
\re Xu H. 1998, PhD Thesis, Shanghai Jiao Tong University
\re Xu H., Ezawa H., Fukazawa Y., Kikuchi K., Makishima K., OhashiT., Tamura T. 
      1997, PASJ 49, 9
\re Xu H., Makishima K., Fukazawa Y., Ikebe Y.,  Kikuchi K., Ohashi T., Tamura, T. 
      1998, ApJ 500, 
\re Yamashita A., Dotani T., Ezuka H., Kawasaki M., Takahashi K. 
      1999, Nucl. Instr. Meth. A436, 68
\re Yan L, Cohen J. G. 1995, ApJ 454, 44
%%%%%%%%%%%%%%%%%%%%%%%%%%%%%%%%%%%%%%%%%%%%%%%%%%%%%%%%%%%%%%%%%%%%%%%%%%%

\onecolumn
\normalsize

%==================   Table 1  =============================================
\begin{table}[b]
\caption{Two-temperature modeling of a sample of nearby clusters,
quoted from Fukazawa (1997) and Fukazawa et al. (2000).}
\begin{center}
\begin{small}
\begin{tabular}{lccccccc}
\hline
\hline
object   &redshift   & B-M$^{*}$ & $T_{\rm h}^{\dagger}$ & $T_{\rm c}^{\dagger}$  
      &$Q_{\rm c}/Q_{\rm h}^{\ddagger}$ &$L_{\rm c}^{\S}$ &$\eta(\%)^{\#}$\\  
\hline
A1795   	&0.062 & I    &$5.89\pm0.15$&$2.04\pm0.63$&$0.22\pm0.07$& 6.3  & $2.6\pm1.4$\\  
A119 	   &0.044 &II-III&$5.52\pm0.27$&     ---     &$  < 0.16   $&$<1.2$& \\ 
A3558 	  &0.048 & I    &$5.17\pm0.25$&     ---     &$  < 0.05   $&$<1.5$& \\
A2147 	  &0.036 & III  &$4.92\pm0.22$&     ---     &$  < 0.20   $&$<0.6$& \\
A496 	   &0.032 & I    &$4.12\pm0.12$&$1.80\pm0.14$&$0.62\pm0.08$& 3.2  & $10.5\pm1.8$\\
A2199 	  &0.030 & I    &$4.10\pm0.07$&$1.78\pm0.32$&$0.17\pm0.05$& 1.2  & $3.1\pm1.2$\\
A4059   	&0.048 & I    &$4.03\pm0.11$&$1.38\pm0.25$&$0.08\pm0.06$& 0.5  & $0.9\pm0.7$\\
AWM7    	&0.018 & I    &$3.76\pm0.08$&$1.83\pm0.46$&$0.27\pm0.13$& 0.3  & $6.0\pm3.5$\\
Centaurus&0.011 & I-II &$3.75\pm0.07$&$1.37\pm0.02$&$1.02\pm0.09$& 0.8  & $12.4\pm1.2$\\
MKW3s 	  &0.043 &II-III&$3.69\pm0.20$&$1.45\pm0.61$&$0.09\pm0.07$& 0.6  & $1.4\pm1.4$\\
A2063 	  &0.034 & II   &$3.68\pm0.11$&$1.64\pm0.64$&$0.19\pm0.12$& 0.4  & $3.6\pm3.0$\\
A2634 	  &0.031 & II   &$3.64\pm0.26$&$1.45\pm0.66$&  $< 1.19$   &$<0.3$&  \\
Hydra-A 	&0.052 & --   &$3.59\pm0.10$&$2.15\pm0.90$&$0.28\pm0.24$& 3.7  & $9.2\pm9.5$\\
A1060 	  &0.011 & III  &$3.25\pm0.05$&     ---     &$  < 0.04   $&$<0.1$& \\
A539 	   &0.027 & III  &$3.21\pm0.08$&     ---     &$  < 0.19   $&$<0.1$&\\
3A0335+096&0.035& --   &$3.03\pm0.07$&$1.42\pm0.07$&$0.65\pm0.09$& 8.0  & $12.6\pm1.9$\\
Virgo   	&0.004 & III  &$2.58\pm0.03$&$1.25\pm0.02$&$1.44\pm0.12$& 0.5  & $25.2\pm2.3$\\
AWM4   	 &0.031 & I    &$2.38\pm0.17$&$1.14\pm0.23$&$< 0.22     $&$<0.2$&\\
A400 	   &0.023 &II-III&$2.33\pm0.14$&     ---     &$ <0.13     $&$<0.1$& \\
A262 	   &0.016 & III  &$2.20\pm0.04$&$1.11\pm0.07$&$0.25\pm0.19$& 0.3  & $6.3\pm4.2$\\
\hline
\end{tabular}
\end{small}
\end{center}
$^{*}$ The Bautz-Morgan morphological type.\\
$^{\dagger}$ Hot-phase temperature and cool-phase temperature, both in keV.\\
$^{\ddagger}$ Ratio of mission integrals for the hot (suffix h) and 
       cool (suffix c) phases, calculated over a central region of 
       $\sim 100 \ h_{50}^{-1}$ kpc, in unit of cm$^{-3}$. \\
$^{\S}$ The 0.5--3 keV cool-phase luminosity in $10^{43}$ ergs cm$^{-2}$ s$^{-1}$.\\
$^{\#}$ The volume filling factor of the cool phase 
        in the central $\sim 100 \ h_{50}^{-1}$ kpc, in units of percent.
\end{table}
%===================   Table 1  ==========================================

\clearpage
\twocolumn
%====================== Figure 1 ========= OK =============================
%=======================================================================

%====================== Figure 2 ======== OK ==============================
%=======================================================================

%====================== Figure 3  ====== OK ================================
%=======================================================================

%====================== Figure 4 ======================================
%=======================================================================

%====================== Figure 5 =====================================
%======================================================================

%====================== Figure 6 ===== OK =================================
%=====================================================================

%====================== Figure 7 ======================================
%======================================================================

%====================== Figure 8 ========= OK =============================
%======================================================================

%====================== Figure 9 ========= OK =============================
%======================================================================

\end{document}